\documentclass{iopjournal}
\usepackage[utf8x]{inputenc}
\usepackage{graphicx}
\usepackage{dcolumn}
\usepackage{bm}
\usepackage{hyperref}

\usepackage{xcolor,soul}

\usepackage{amsmath}
\usepackage{amssymb}

\usepackage{cite}

\newcommand*{\plath}{C^{(2)}_\infty}

\begin{document}
	\raggedbottom

	\articletype{Paper}

	\title{Harnessing finite-size effects to gauge aging in the \texorpdfstring{$2D$}{2D} Ising model}

	\author{Dustin Warkotsch$^{1,*}$\orcid{0009-0009-3081-2468}, Malte Henkel$^{2,3}$\orcid{0000-0002-5048-7852} and Wolfhard Janke$^{1}$\orcid{0000-0002-5165-9097}}

	\affil{$^1$Institut f\"ur Theoretische Physik, Universit\"at Leipzig, IPF 231101, D-04081 Leipzig, Germany}

	\affil{$^2$Laboratoire de Physique et Chimie Th\'eoriques (CNRS UMR 7019), Universit\'e de Lorraine Nancy, B.P. 70239, F - 54506 Vand{\oe}uvre-l\`es-Nancy Cedex, France}

	\affil{$^3$Centro de F\'{\i}sica Te\'orica e Computacional, Universidade de Lisboa, Campo Grande, P–1749-016 Lisbon, Portugal}

	\affil{$^*$Author to whom any correspondence should be addressed.}

	\email{dustin.warkotsch@itp.uni-leipzig.de}


	\begin{abstract}
		The relaxation behavior towards equilibrium of the $2D$ Ising model with nearest-neighbor interactions has been studied with focus on the two-time autocorrelator~$C(t,s)$. Finite-size effects affecting the growing magnetic domains lead to the saturation of $C(t,s)$ with a distinct plateau of height $C_{\infty}^{(2)}(s,L)$ scaling algebraically with waiting time $s$ and lattice size $L$. These scaling relations are used to produce precise estimates for the autocorrelation exponent $\lambda$ and dynamical exponent $z$ with deliberately small lattices. Treating smooth domain walls in a similar manner to the lattice boundaries, their effect on $C(t,s)$ can be understood as \emph{premature} finite-size phenomenon, extending our ansatz to systems not yet in equilibrium.
	\end{abstract}

\setcounter{footnote}{3}

	%%%%%%%%%%%%%%%%%%%%%%%%%%%%%%%%%%%%%%%%%%%%%%%%%%%%%%%%%%%%%%%%%%%%%%%%%%%%%%%%%%%%%%%%%%%%%%%%%%%%%%%%%%%%%%%%%%%%%%%%%%%%%
	\section{\label{sec:introduction}Introduction}
	%%%%%%%%%%%%%%%%%%%%%%%%%%%%%%%%%%%%%%%%%%%%%%%%%%%%%%%%%%%%%%%%%%%%%%%%%%%%%%%%%%%%%%%%%%%%%%%%%%%%%%%%%%%%%%%%%%%%%%%%%%%%%

	Relaxation phenomena far from equilibrium continue to pose many challenging questions.
	An important sub-class are {\em physical aging phenomena},
	which arise when a system, after an initial preparation in some state which is almost always taken as a totally disordered initial state,
	is quenched to a temperature $T$ which is either onto the critical temperature $T_c>0$ or else into the phase-coexistence region $T<T_c$
	where at least two physically distinct thermodynamical states exist~\cite{struikPhysicalAgingPlastics1977,struikPhysicalAgingAmorphous1978a,brayTheoryPhaseorderingKinetics2002,tauberCriticalDynamics2014,henkelAgeingDynamicalScaling2010,puriKineticsPhaseTransitions2009,mazenkoNonequilibriumStatisticalMechanics2008,cugliandoloDynamicsGlassySystems2002}.
	The phase-transition at $T_c$ is described by a physical order parameter $\phi$ (in our study identified with the
	magnetization).
	In what follows we shall restrict to quenches into the ordered phase with $T<T_c$
	and if the dynamics is such that the order parameter is not conserved, one speaks of {\em phase-ordering kinetics} \cite{brayGrowthLawsPhase1994}.
	Microscopically, a spatially infinite system is inhomogeneous, as it locally decomposes into ordered clusters whose linear size $\ell(t)$
	grows with time.
	If the ordering kinetics is such that one finds an algebraic law $\ell(t)\sim t^{1/z}$ for large enough times, this defines the {\em dynamical exponent} $z$.
	The competition between the different equilibrium states leads to a slow-down of the kinetics such that the aging processes can be characterized phenomenologically by the three
	defining properties~\cite{henkelAgeingDynamicalScaling2010,struikPhysicalAgingPlastics1977,tauberCriticalDynamics2014}:
	\begin{enumerate}
		\item Slow dynamics, i.e., slower than described by a simple exponential.
		\item Absence of time-translation invariance.
		\item Dynamical scaling.
	\end{enumerate}
	A useful characterization of the aging process can be given through studying the unconnected two-time autocorrelator, defined as
	\begin{equation} \label{eq:two-time-autocorrelation}
		C(t,s) = \left\langle \phi(t,\vec{r}) \phi(s,\vec{r}) \right\rangle
	\end{equation}
	where $\phi=\phi(t,\vec{r})$ is the coarse-grained local order parameter depending on time $t$ and the spatial coordinate $\vec{r}$.
	Furthermore $s$ denotes the {\em waiting time} and $t>s$ the {\em observation time}.
	In phase ordering, this autocorrelator assumes for sufficiently long times a scaling form
	\begin{equation} \label{eq:dynamical_scaling}
		C(t,s) = f_C\left(\frac{t}{s}\right) \;\; , \;\; f_C(y) \stackrel{y\gg 1}{\simeq} y^{-\lambda/z}
	\end{equation}
	with the expected long-time asymptotics for $y=t/s\gg 1$. Herein, $\lambda$ is the {\em autocorrelation exponent} which is independent of the
	equilibrium critical exponents~\cite{fisherNonequilibriumDynamicsSpin1988}.
	After the verification of the scaling (\ref{eq:dynamical_scaling}),
	the determination of exponents such as $z$ or $\lambda$ is a central task in theoretical studies of phase-ordering kinetics.

\begin{figure}[t]
		\centering
		\includegraphics[width=\linewidth]{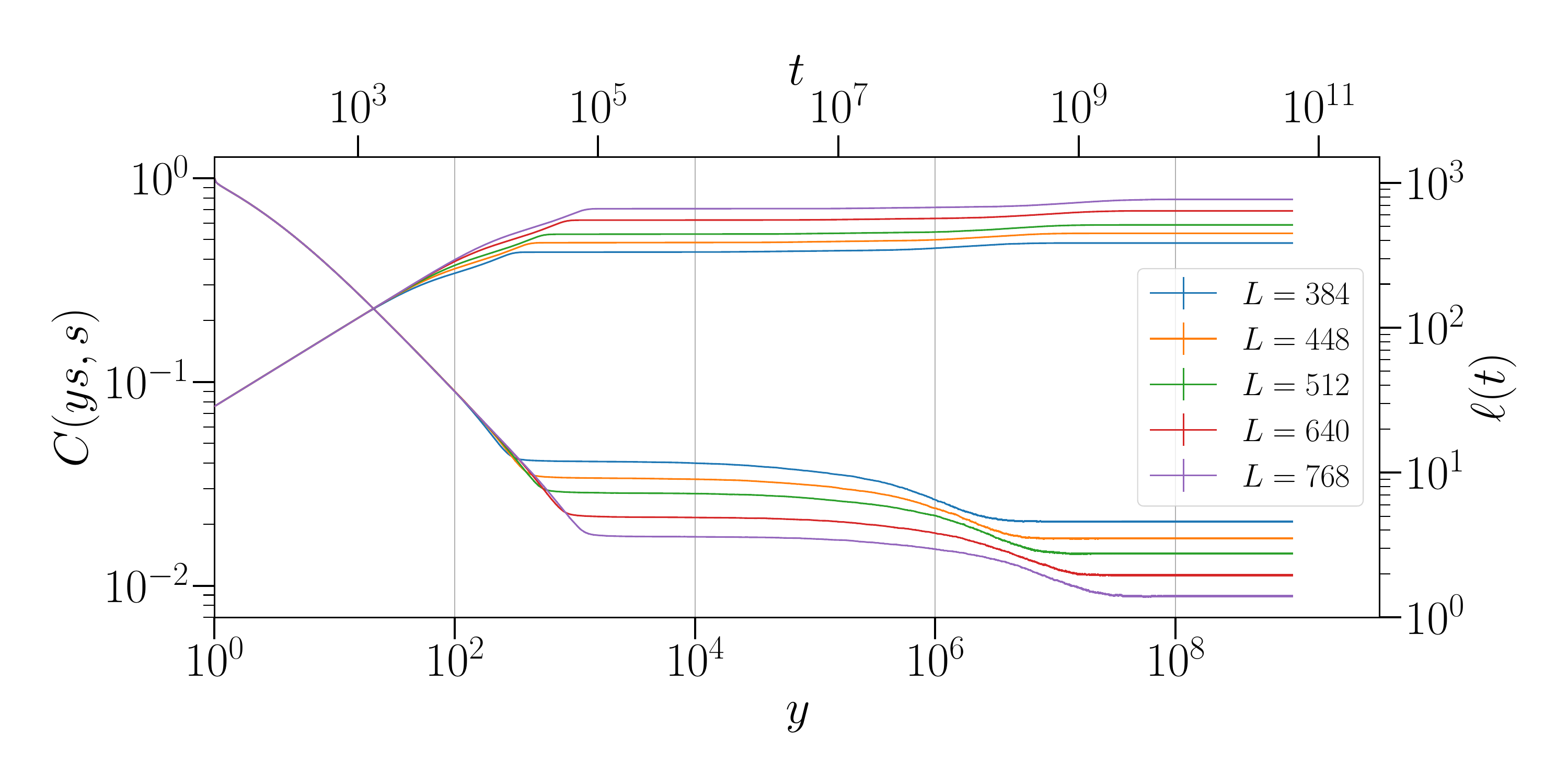}
		\caption[cvsl]{Exemplary behavior of the characteristic length scale $\ell(t)$ vs time $t$ (upper curves, right and upper axes) and the two-time autocorrelator $C(ys,s)$ vs $y$ (lower curves, left and lower axes) for a nearest-neighbor $2D$ Ising model after a quench to $0.2T_c$ for system sizes $L=\{384, 448, 512, 640, 768\}$ and waiting time $s=64$. Note that saturation of both ex\-hib\-its a step-wise behavior featuring the formation of an initial pseudo-plateau before settling at the equilibrium value beginning around $t \approx 10^{10}$ and $y \approx 10^8$, respectively.}
		\label{fig:cvsl}
	\end{figure}

	The above phenomenology refers to  spatially infinite systems.
	Since any lattice simulation works with fully finite systems, or any experiment should count with a certain granularity of the sample,
	it is necessary to analyze what happens in systems of linear size $L$.
	Two main effects are shown in figure~\ref{fig:cvsl}.
	%============================================================================================================================%%
	%============================================================================================================================%%
	The upper set of curves shows the characteristic length scale $\ell(t)$ of the microscopic clusters.
	Here, $\ell(t)$ is obtained via the mean $\langle l \rangle$ of the histogram of domain lengths $P(l,t)$ with each individual $l$ being measured as the distance between two successive domain boundaries along the horizontal and vertical cross sections of the lattice (while minding the periodic boundary conditions)~\cite{dasDynamicsPhaseSeparation2002,midyaAgingFerromagneticOrdering2014,majumderDomainCoarseningTwo2010,tafaKineticsDomainGrowth2001}.
	For $t$ not too large, the finite extent of the system is not yet noticeable and $\ell(t)\sim t^{1/z}$ increases as a power law.
	As time passes, $\ell(t)$ becomes comparable to $L$ in magnitude and saturates on the finite lattice.
	A similar behavior is also reflected in the two-time autocorrelator $C(ys,s)$, as seen in the lower set of curves.
	Data are chosen such that $\ell(s)$ is not yet in the saturation regime but past the initial transient.
	When $y=t/s$ is not yet too large, the autocorrelator is very close to the one of the spatial infinite system which is apparent from the initial collapse of the data for different lattice sizes $L$ in figure~\ref{fig:cvsl}.
	However, when $y$ is increased, the correlator first decreases faster than it would be the case for a spatially infinite system -- a behavior also observed in the spherical model (see especially figure~1(b) of reference~\cite{henkelNonequilibriumRelaxationsAgeing2023}), but not in the $1D$ Glauber-Ising model at $T=T_c=0$~\cite{henkelFinitesizeScalingAgeing2025} -- before saturating to a characteristic plateau.
	Specifically regarding the $2D$ Ising system investigated here, this saturation has been found to take place in multiple steps: First, a \emph{pseudo-plateau} emerges which mimics the behavior in equilibrium at $t \rightarrow \infty$.
	However, the system is not yet fully equilibrated but rather stuck in metastable states featuring striped configurations, whose domain walls limit the characteristic length scale $\ell(t)$ similarly as the system boundary itself.
	As the lattice stays en route towards equilibrium, these striped states gradually vanish, such that the \emph{actual} saturation is reached, yielding a concrete plateau at the far end of figure~\ref{fig:cvsl}.

	The cross-over from infinite system behavior to saturation is seen to occur at the same time when the characteristic length changes from the power-law growth $\ell(t)=\ell(ys)\sim \bigl( ys\bigr)^{1/z}$ towards saturation.
	We see that the characteristic length scale indeed shows for not too large times the infinite-size power law but crosses over to saturation when now $\ell\sim L$ for larger times, as is well-known \cite{christiansenAgingLongrangeIsing2020}.
	Similarly, the autocorrelator $C(ys,s)$ obeys the $s$-independent scaling (\ref{eq:dynamical_scaling}) with the expected power-law decay.
	But for the same values of $y$ where $\ell(t)$ is seen to saturate at $L$, we also observe that $C$ crosses over to a plateau, whose height
	\begin{equation}
		\label{eq:C_inf_definition}
		C_{\infty}^{(2)} = C_{\infty}^{(2)}(s;L) := \lim_{y\to\infty} C(ys,s;L)
	\end{equation}
	is decreasing with $L$.

	Heuristically, these observations from figure~\ref{fig:cvsl} can be understood as follows, using a finite-size scaling ansatz.
	Since for $y$ large enough, and with (\ref{eq:dynamical_scaling}), one has in infinite volume, i.e., $L \to \infty$
	\begin{equation} \label{eq:total-scaling}
		C(t,s) \sim \left( \frac{\ell(t)}{\ell(s)} \right)^{-\lambda} .
	\end{equation}
	Now if one takes the observation time $t$ large enough that $\ell(t)\sim L$ has saturated whereas	the waiting time $s$ is kept large enough that the power law~(\ref{eq:dynamical_scaling}) is observed but yet small enough that saturation in $\ell(s)$ not yet occurs,
	the autocorrelator will become constant in $y=t/s$ with
	a plateau height
	\begin{equation} \label{eq:both-scalings}
		\plath\sim \bigl( L s^{-1/z} \bigr)^{-\lambda}~.
	\end{equation}
	Thus, if one keeps $s$ fixed, this implies
	\begin{equation} \label{eq:L-scaling}
		\plath \sim L^{-\lambda}
	\end{equation}
	and if one keeps $L$ fixed instead, one should have
	\begin{equation} \label{eq:s-scaling}
		\plath \sim s^{\lambda/z} ~.
	\end{equation}

	The main interest of these results is that they offer a new computational access for finding values of $\lambda$ and $\lambda/z$.
	Specifically, they provide a new way of estimating $\lambda$ itself, independently of $z$.
	Furthermore, this work addresses the question if the predictions (\ref{eq:L-scaling}),~(\ref{eq:s-scaling}) can furnish an efficient computational tool. As we shall see, there are indeed important technical questions to address which turn out to be far from being straightforward.

	Formally, the predictions (\ref{eq:L-scaling}),~(\ref{eq:s-scaling}) can be derived in a novel general approach to aging, based on generalized time-translation invariance~\cite{henkelGeneralisedTimetranslationinvarianceSimple2025,henkelPhysicalAgeingGeneralised2025}.
	Analytical confirmations of the scaling laws (\ref{eq:L-scaling}),~(\ref{eq:s-scaling}) exist in the spherical model for $2<d<4$ dimensions and quenched to $T<T_c$ \cite{henkelNonequilibriumRelaxationsAgeing2023}, the $1D$ Glauber-Ising model quenched to $T=0$ \cite{henkelFinitesizeScalingAgeing2025} and the mean-field $p=2$
	spherical spin glass quenched to $T=0$ \cite{fyodorovLargeTimeZero2015}.
	Here we present the first investigation in a purely numerical study and shall concentrate on the emerging issues when trying to apply
	equations~(\ref{eq:L-scaling}),~(\ref{eq:s-scaling}) to simulational data.
	Furthermore, this thorough analysis may prove itself advantageous for the application on experimental data as well.
	More specifically, recent coarsening experiments with $2D$ nematic liquid crystals~\cite{almeidaPhaseorderingKineticsAllenCahn2021} have demonstrated dynamics and scaling behavior comparable to a spin system with nonconserved order parameter.
	Thus, we expect our analysis via the $2D$ Ising model to provide insights regarding the application to real-world data in a similar manner.

	It is well-established that in phase-ordering kinetics of systems with short-range interactions and a nonconserved order parameter, the dynamical exponent $z=2$~\cite{brayGrowthLawsPhase1994,rutenbergEnergyscalingApproachPhaseordering1995} which has also been tested extensively in the literature, see \cite{henkelAgeingDynamicalScaling2010} and references therein.
	The value of $\lambda$ on the other hand remains contentious.
	Through general scaling arguments, Fisher and Huse established a lower bound yielding $\lambda\geq d/2$~\cite{fisherNonequilibriumDynamicsSpin1988}.
	Later, Yeung et al. recovered this relation in a more general sense through the Cauchy-Schwartz inequality as a special case of $\lambda\geq (\beta + d)/2$ for systems with conserved, non-conserved (where $\beta=0$), scalar and vector order parameter~\cite{yeungBoundsDecayAutocorrelation1996}.
	Additionally, Fisher and Huse~\cite{fisherNonequilibriumDynamicsSpin1988} proposed an upper bound $\lambda \leq 5/4$ for the $2D$ Ising universality class.
	However, the arguments used to derive this are certainly less rigorous than those for the lower bound.
	There even exists the conjecture $\lambda \stackrel{?}{=} 5/4$~\cite{fisherNonequilibriumDynamicsSpin1988}.
	While it is compatible with several numerical simulations~\cite{christiansenAgingLongrangeIsing2020,lorenzNumericalTestsLocal2007,henkelTwotimeAutocorrelationFunction2004,menyhardDomaingrowthPropertiesTwodimensional1994} and experiments~\cite{masonScalingBehaviorTwotime1993,almeidaPhaseorderingKineticsAllenCahn2021}, there exist also
	contradicting results, analytical \cite{mazenkoPerturbationExpansionPhaseordering1998} which give $\lambda \approx 1.29$, or numerical \cite{midyaAgingFerromagneticOrdering2014} who find $\lambda \approx 1.32$.

	An alternative and independent way for determining $\lambda$ or $\lambda/z$ would therefore be of value.
	At least, a new computational technique should not suffer from the same biases as the older ones.
	In principle, in using equations~(\ref{eq:L-scaling}),~(\ref{eq:s-scaling}) it might be sufficient to work with relatively small lattices since we do not try to avoid finite-size effects at all costs but rather try to harness and exploit them.
	This contrasts with the usual necessity of continuously pursuing larger systems in order to increase the usable fitting window of $C(ys,s)$.
	Besides, since we shall obtain separate estimates for $\lambda$ and $\lambda/z$, as a by-product we can also submit the long-standing theoretical expectation $z=2$~\cite{rutenbergEnergyscalingApproachPhaseordering1995, brayGrowthLawsPhase1994} to a new kind of test.

	The rest of the paper is organized as follows.
	In section~\ref{sec:model_and_simulational_methods}, we present our model as well as simulational techniques, algorithms and parameters.
	The analysis of our data is described through section~\ref{sec:data_analysis} in great detail.
	After having introduced our methodology, we present our results in section~\ref{sec:results}.
	More precisely, a large part of our work is dedicated to uncover systematic biases in our finite-size method and their rectification.
	This is done in subsection~\ref{subsec:0.2Tc} which also leads to the results for $T=0.2Tc$ where we present our estimates regarding $\lambda$ and ${\lambda}/{z}$.
	The specifics and prominence of the challenges mentioned in subsection~\ref{subsec:0.2Tc} for $T=0.2T_c$ are highlighted in subsection~\ref{subsec:0.1Tc} for the lower temperature $T=0.1T_c$ where several distinct possibilities are shown to overcome the emerging problems.
	Finally, the paper closes with our conclusions and a discussion of additional ways to utilize our method and apply it to distinct models in section~\ref{sec:conclusion}.

	%%%%%%%%%%%%%%%%%%%%%%%%%%%%%%%%%%%%%%%%%%%%%%%%%%%%%%%%%%%%%%%%%%%%%%%%%%%%%%%%%%%%%%%%%%%%%%%%%%%%%%%%%%%%%%%%%%%%%%%%%%%%%
	\section{\label{sec:model_and_simulational_methods}Model and simulational methods}
	%%%%%%%%%%%%%%%%%%%%%%%%%%%%%%%%%%%%%%%%%%%%%%%%%%%%%%%%%%%%%%%%%%%%%%%%%%%%%%%%%%%%%%%%%%%%%%%%%%%%%%%%%%%%%%%%%%%%%%%%%%%%%

	In order to systematically examine the practicality of our method of estimation, we investigate the Ising model with nonconserved order parameter driven by nearest-neighbor interactions on a square periodic lattice with $N=L^2$ spins, using Markov chain Monte Carlo simulations~\cite{landauGuideMonteCarlo2009}.
	This model is the simplest non-trivial framework in which this method should be numerically exploitable after it has been proven for the spherical model~\cite{henkelNonequilibriumRelaxationsAgeing2023}.
	Its Hamiltonian is given by:
	\begin{equation}
		\label{eq:Hamiltonian}
		\mathcal H = -J \sum_{\langle ij \rangle} \sigma_i \sigma_j
	\end{equation}
	with coupling constant~$J=1$ between the neighboring spins~$\sigma=\pm 1$ at sites~$i$ and $j$ implementing ferromagnetic interactions.
	Here, $\langle ij \rangle$ symbolizes summation over all such neighboring pairs in the whole lattice.
	The main observable in this work -- the two-time autocorrelator $C(t,s)$ -- as given by equation~(\ref{eq:two-time-autocorrelation}) in the most general form, is implemented here as

	\begin{equation}
		C(t,s) = \frac{1}{N} \sum_{i=1}^{N} \sigma_i(t) \sigma_i(s) ~.
		\label{eq:C(t,s)_Ising}
	\end{equation}

	Since the behavior of the two-time autocorrelator is inherently linked to the magnetic domains, the system has to be brought out of equilibrium to study its evolution and the behavior under finite-size constraints.
	This principle is realized by initializing the lattice with a fully random spin configuration retaining a minuscule order parameter, here: magnetization $m$, representing a $T \rightarrow \infty$ sample, before quenching the system well below $T_c \approx 2.27 J/k_B > 0$, with $k_B$ being the Boltzmann constant set to unity here.
	After the quench, the system's evolution is driven forward by spin-flip dynamics following the kinetic Monte Carlo (or $n$-fold way) algorithm~\cite{bortzNewAlgorithmMonte1975,novotnyNewApproachOld1995} with Metropolis-like spin-flip probabilities~\cite{metropolisEquationStateCalculations1953} to maximize efficiency at low temperatures.
	The simulation is carried out with $I=30\,000$ distinct initializations for each lattice size~$L$ realized using different initial configurations and seeds for the pseudo random number generator.

	%%%%%%%%%%%%%%%%%%%%%%%%%%%%%%%%%%%%%%%%%%%%%%%%%%%%%%%%%%%%%%%%%%%%%%%%%%%%%%%%%%%%%%%%%%%%%%%%%%%%%%%%%%%%%%%%%%%%%%%%%%%%%
	\section{\label{sec:data_analysis} Data analysis}
	%%%%%%%%%%%%%%%%%%%%%%%%%%%%%%%%%%%%%%%%%%%%%%%%%%%%%%%%%%%%%%%%%%%%%%%%%%%%%%%%%%%%%%%%%%%%%%%%%%%%%%%%%%%%%%%%%%%%%%%%%%%%%

\begin{figure}[t]
		\centering
		\includegraphics[width=0.6\linewidth]{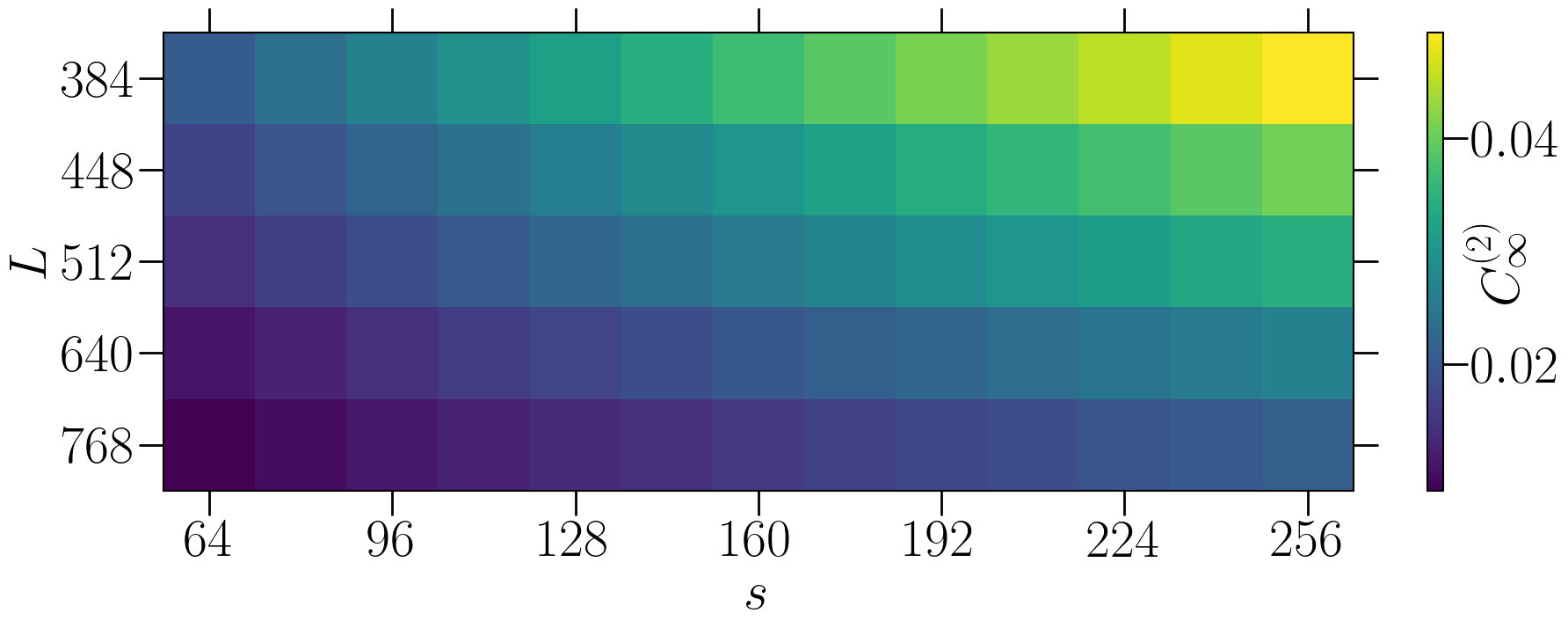}
		\caption[cvsl]{Heatmap of plateau heights $\plath$ of the two-time autocorrelator at $T=0.2T_c$ depending on linear lattice size $L$ and waiting time $s$. As the plateau height grows from the lower left to the upper right side, equations~(\ref{eq:L-scaling}) and~(\ref{eq:s-scaling}) describe the relation between individual values of $\plath$ for their respective orthogonal axis.}
		\label{fig:heatmap}
	\end{figure}

	For each of the distinctly initialized runs, the plateau heights $C^{(2)}_{\infty}$ for each parameter set $(s_i,L_j)$ is taken as the last data point of the respective time series.
	Plotting these data points against both parameters simultaneously, as seen in the heat map of figure~\ref{fig:heatmap}, illustrates the general trend of relations~(\ref{eq:L-scaling}) and~(\ref{eq:s-scaling}), i.e., $C^{(2)}_{\infty}$ grows for smaller $L$ and larger $s$.
	Interestingly, measuring both exponents precisely, i.e., autocorrelation exponent $\lambda$ and dynamical exponent $z$, requires two, in a sense, `orthogonal' fits of these data: equation~(\ref{eq:L-scaling}) fits $C^{(2)}_{\infty}$ against $L$ with fixed $s$, whereas equation~(\ref{eq:s-scaling}) does the opposite; which can be visualized as fitting the two orthogonal axes of the heat map in figure~\ref{fig:heatmap} for each column and line respectively.

	%============================================================================================================================%%
	%============================================================================================================================%%

	Although this procedure sounds simple enough, it is significantly impeded by strong correlations between the $C^{(2)}_{\infty}$ for different waiting times $s$, where the Pearson correlation coefficient $\rho$ can reach a magnitude as high as $\rho \sim 0.999$ between neighboring $s$.
	These correlations stem from the compounded way of measuring at different waiting times, which is done by taking `snapshots' of the lattice at the respective $s$ before measuring the two-time autocorrelation $C(t,s)$ by `comparing' the state at time $t$ to the snapshot taken at time $s$ via equation~(\ref{eq:C(t,s)_Ising}).
	Thereby, all of the various $s$ are taken from the very same run, in order to curb the number of simulations, which leads, however, to inevitable correlations between those snapshots as each state at time $s_2$ directly evolves from the state at time $s_1 < s_2$ through the Markov chain.
	If ignored, these correlations can lead to fairly unpredictable results, ranging from inaccurate measurements to over- or underestimation of errors.
	To solve these issues, there already exist several sufficiently sophisticated methods, which we have to include at different points of our analysis, since only one `fitting axis' (see figure~\ref{fig:heatmap}) is affected by these correlations.

	Thus, we apply the following steps after having taken an estimate of $C^{(2)}_{\infty}$ at each $s$ and $L$ for every run:
	\begin{enumerate}
		\item First, we resample our $30\,000$ runs of $C^{(2)}_{\infty}$ for each $s$ and $L$ with the Jackknifing technique~\cite{efronJackknifeBootstrapOther1982,jankeMonteCarloSimulations2012,weigelErrorEstimationReduction2010}, such that we are left with $J=300$ \emph{blocks} each containing all runs minus a distinct $100$.
		From each block, we retrieve an average and an error estimate, which we can use for an overall error estimation as well as correlational analysis in step 3.
		\item To retrieve an estimate $\lambda(s)$ for a single fixed waiting time $s$, we perform an \emph{error-weighted} fit of $C^{(2)}_{\infty}$ against $L$ using equation~(\ref{eq:L-scaling}) (see figure~\ref{fig:withfit_0.2Tc}(a) below) for each of the $300$ Jackknife blocks and average over those.
		This procedure is then repeated for all $s \in \{64, 96, 128, 160, 192, 256\}$ which we investigate here.
		\item It should be noted that each fit using equation~(\ref{eq:L-scaling}) is in itself a valid estimator since the plateau heights do not exhibit correlations along the $L$-axis.
		However, for a total average over all fixed $s$, they need to be taken into account.
		This can be done, using the \emph{correlation-weighted average}~\cite{weigelErrorEstimationReduction2010,weigelCrossCorrelationsScaling2009} directly building on Jackknife blocks.
		Thereby, we can estimate the covariance matrix
		\begin{equation*}
			\hat{\Gamma}_{ij}=\frac{J-1}{J} \sum_{k=1}^J \left[\lambda_k(s_i) - \lambda(s_i)\right] \left[\lambda_k(s_j) - \lambda(s_j)\right]
		\end{equation*}
		of our $\lambda(s_i)$ estimates at $s_i$ over all Jackknife blocks with $\lambda_k(s_i)$ being the $\lambda$ estimate at $s_i$ of the $k$-th Jackknife block.
		This then leads to the optimal weights\,\footnote{See reference~\cite{weigelErrorEstimationReduction2010} equation~(22)} for averaging all $\lambda(s)$ with the minimal variance, resulting in the final estimate $\lambda$.

		\item To retrieve an estimate $(\lambda/z)(L)$ for a single fixed lattice size $L \in \{384, 448, 512, 640, 768\}$, it is necessary to perform a \emph{correlated} fit~\cite{michaelFittingCorrelatedData1994,andersonIntroductionMultivariateStatistical2003,brunoFitsCorrelatedAutocorrelated2023} of $C^{(2)}_{\infty}$ against $s$ using equation~(\ref{eq:s-scaling}) (see figure~\ref{fig:withfit_0.2Tc}(b) below).
		This is done with all runs at once as Jackknifing is not necessary.
		More specifically, this method hinges on the accuracy of the covariance matrix
		\begin{equation*}
			\begin{split}
				\hat{\Gamma}_{ij}(L)=\frac{1}{I-1} \sum_{k=1}^I & \left[C_{\infty,k}^{(2)}(s_i,L) - C_{\infty}^{(2)}(s_i,L)\right] \times \\
				& \times \left[C_{\infty,k}^{(2)}(s_j,L) - C_{\infty}^{(2)}(s_j,L)\right],
			\end{split}
		\end{equation*}
		used to weight each plateau height $\plath(s_i,L)$ according to its uncertainty together with the covariance between it and all other data points at the same $L$.
		Here, $C_{\infty,k}^{(2)}(s_i,L)$ is the plateau height measured at $s_i$ and $L$ in the $k$-th initialization, and $C_{\infty}^{(2)}(s_i,L)$ is the average plateau height at $s_i$ and $L$ over all $I$ initializations.
		In our work, $\hat{\Gamma}_{ij}(L)$ can be precisely estimated using $I=30\,000$ runs for each $C^{(2)}_{\infty}$.

		Then, similarly to an error-weighted fit, one estimates the regression coefficients $A$ by minimizing the distance between model $\Phi(A)$ and data $Y$ (here, the plateau heights) weighted by the inverse covariance matrix $\hat{\Gamma}^{-1}$~\cite{brunoFitsCorrelatedAutocorrelated2023}:
		\begin{equation*}
			\chi^2(A) = (Y - \Phi(A))^T \hat{\Gamma}^{-1}(L) (Y - \Phi(A)) .
		\end{equation*}

		\item Since the resulting $(\lambda/z)(L)$ for all $L$ are uncorrelated with each other, they can be combined using an error-weighted average to form the final estimate $\lambda/z$.

	\end{enumerate}

	%%%%%%%%%%%%%%%%%%%%%%%%%%%%%%%%%%%%%%%%%%%%%%%%%%%%%%%%%%%%%%%%%%%%%%%%%%%%%%%%%%%%%%%%%%%%%%%%%%%%%%%%%%%%%%%%%%%%%%%%%%%%%
	\section{\label{sec:results}Results}
	%%%%%%%%%%%%%%%%%%%%%%%%%%%%%%%%%%%%%%%%%%%%%%%%%%%%%%%%%%%%%%%%%%%%%%%%%%%%%%%%%%%%%%%%%%%%%%%%%%%%%%%%%%%%%%%%%%%%%%%%%%%%%

	\subsection{\label{subsec:0.2Tc}Estimates at \texorpdfstring{$T=0.2T_c$}{T=0.2Tc}}

	In this subsection, we explore the straightforward implementation of the procedure described in section~\ref{sec:introduction}, including the detailed path from initial measurements to our exponents' estimation as outlined in section~\ref{sec:data_analysis}.
	To begin, we must choose where in the plateaus (\ref{eq:C_inf_definition}) of the two-time autocorrelation we expect the equilibrium state to have been reached, such that enough time has passed for $y = t/s \gg 1$ to be fulfilled.

	The necessity of a careful choice is illustrated via figure~\ref{fig:c-tsl768_0,2Tc_full_equilibration}, showing the variation of $C(ys,s)$ at $T=0.2T_c$ for the waiting time $s=256$ (figure~\ref{fig:c-tsl768_0,2Tc_full_equilibration}(a)) respectively the lattice size $L=768$ (figure~\ref{fig:c-tsl768_0,2Tc_full_equilibration}(b)) with the other parameter varying after performing $t=6\times10^{10}$ sweeps.
	Noticeably, the approach towards equilibrium produces \emph{two} distinct plateaus, and not a single one as expected.
	Per equation~(\ref{eq:total-scaling}), there ought to be a connection to the characteristic length scale $\ell(t)$ whose growth mirrors the decrease in $C(ys,s)$.
	Thus, the first \emph{pseudo}-plateau which can be identified at $y \approx 10^4 \dots 10^5$, can be interpreted as sign of \emph{premature finite-size effects} temporarily limiting the domains' growth\,\footnote{One may assume that earlier studies which observed `finite-size effects' at large $t$ have only encountered this pseudo-plateau.}.
	As these effects apparently yield a strong influence on the autocorrelator, they may prove themselves problematic for the application of our relations such that a diligent analysis is inevitable, which is a large part of this work.

	%============================================================================================================================%%
	\begin{figure}[t]
		\centering
		\includegraphics[width=1.0\linewidth]{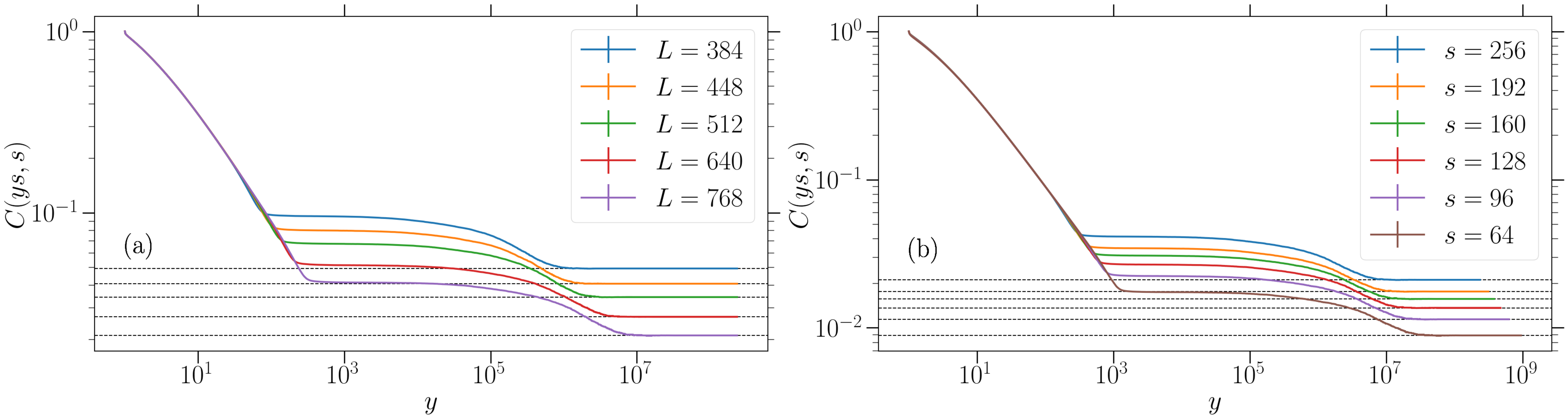}
		\caption{Two-time autocorrelator $C(ys,s)$ for the $2D$ Ising model quenched to $T=0.2T_c$ for various fixed (a) lattice sizes $L$ at $s=256$, or (b) waiting times $s$ at $L=768$.
			The steps in the plateaus result from gradual evaporation of metastable, i.e., striped states of all kinds (viz. figure~\ref{fig:stripes}). The dashed black lines indicate the height $\plath$ for the respective $L$ or $s$.}
		\label{fig:c-tsl768_0,2Tc_full_equilibration}
	\end{figure}
	%============================================================================================================================%%
	%============================================================================================================================%%
	\begin{figure}[t]
		\centering
		\includegraphics[width=0.15\linewidth]{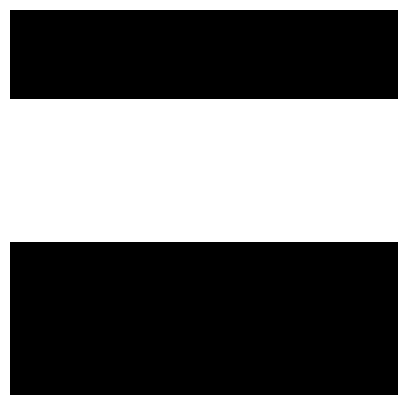}
		\vspace{0.5cm}
		\includegraphics[width=0.15\linewidth]{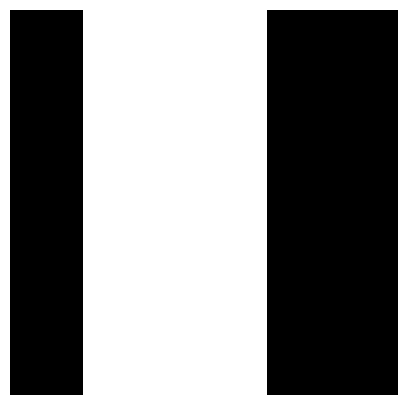}
		\vspace{0.5cm}
		\includegraphics[width=0.15\linewidth]{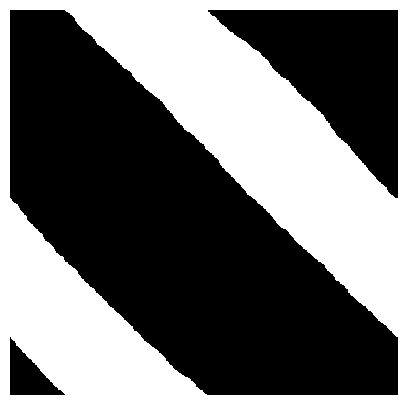}
		\caption{Metastable states of the $2D$ Ising model with periodic boundary conditions:
			horizontal stripe, vertical stripe and diagonal stripe.	The former two are infinitely long-lived at $T=0$
			whereas the lifetime of the last one depends on lattice size $L$. Above $T=0$, all stripes evaporate eventually, at sufficiently large times.}
		\label{fig:stripes}
	\end{figure}
	%============================================================================================================================%%

	In order to study the validity of (\ref{eq:L-scaling}) or (\ref{eq:s-scaling}) we must require that $\ell(t)\lesssim L$ saturates, or equivalently that the observation time (measured in lattice sweeps) $t\gg L^2$~\cite{henkelNonequilibriumRelaxationsAgeing2023}.
	The consequences are, thus, twofold: (a) $L$ should be small enough to display the fully-finite size regime;
	(b) the simulation time must be large enough for the system (at all lattice sizes) to reach equilibrium.
	Satisfying these requirements is not always easy: In the Ising model, a non-negligible portion of initializations may evolve towards some metastable state.
	Specifically in $2D$, examples are given by certain striped configurations, see figure~\ref{fig:stripes}.

\begin{figure}[t]
		\centering
		\includegraphics[width=0.7\linewidth]{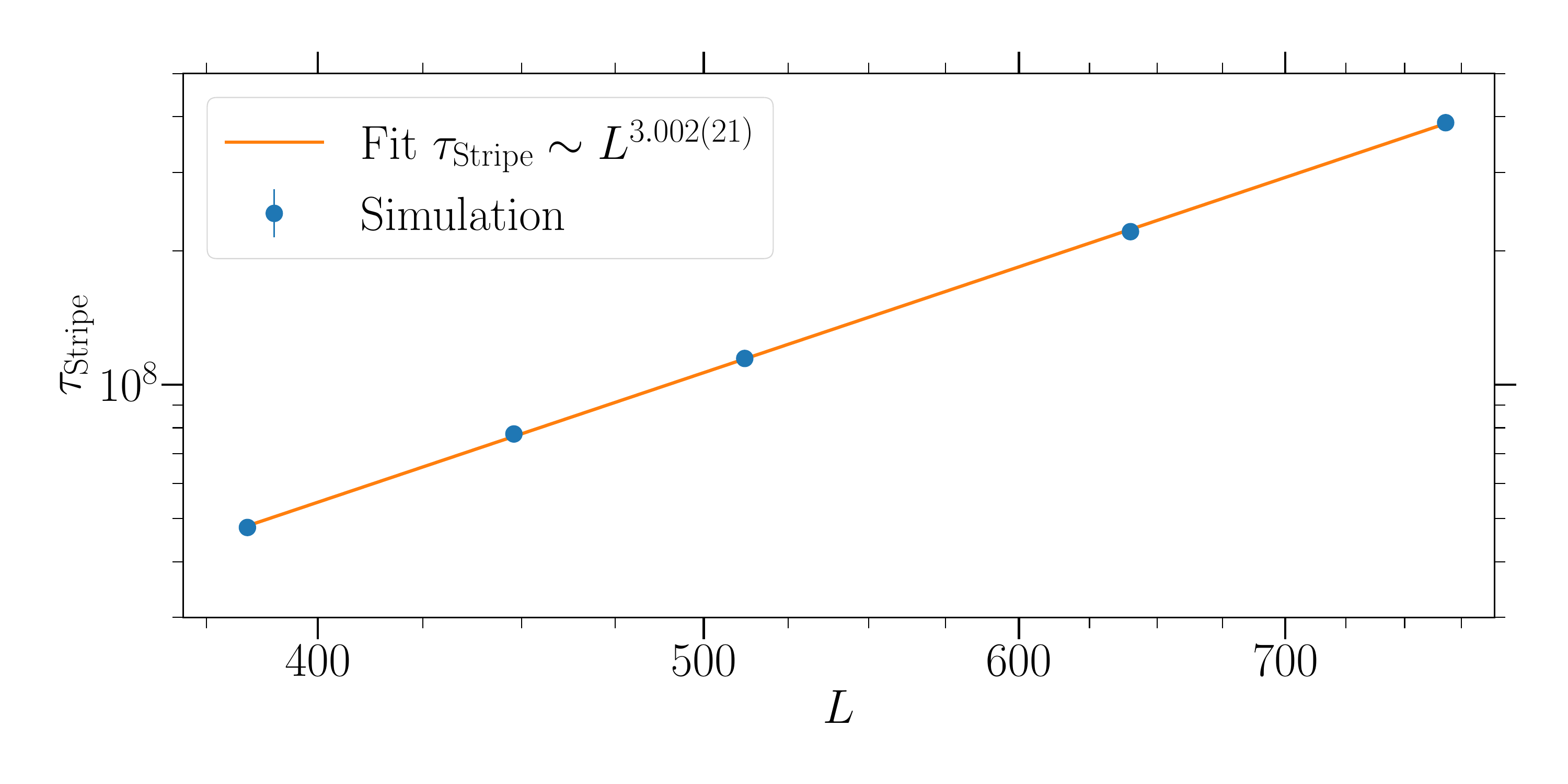}
		\caption{Average lifetime (measured in MC sweeps) of all striped configurations vs lattice size $L$ at $T=0.2T_c$ in $2D$.}
		\label{fig:diagonal_stripes_tau_vs_L}
	\end{figure}

	In fact, it is known that at $T=0$ on a square Ising lattice with short range interactions and periodic boundary conditions, around a third of systems reach \emph{infinitely} long-lived metastable states with smooth stripes in vertical or horizontal direction~\cite{barrosFreezingStripeStates2009,olejarzFate2DKinetic2012}, while only $62\,\%$ of all runs are expected to reach full alignment.
	This behavior is specific to the two-dimensional Ising systems since for $d=1$, equilibrium is always reached owing to geometry, whereas at $d \geq 3$ this almost never occurs.
	Furthermore, in $2D$ systems a small fraction of $\sim 4\,\%$, yields diagonal stripes with a finite life-time $\sim L^3$ at \emph{all} temperatures\,\footnote{More precisely, the ratio of stripes percolating in only one direction is $33.88\,\%$, diagonal stripes happen with a probability of $4.196\,\%$ such that percolation in both directions, i.e., effectively and eventually full equilibration occurs with $61.924\,\%$ chance. These numbers, however, depend significantly on $L$ and $T$~\cite{olejarzFate2DKinetic2012}.}.
	Above absolute zero, this also applies to the other stripes which only percolate in one direction, with a lifetime of
	$\tau_\mathrm{Stripe} \sim e^{4J/k_BT} L^{3}$, where the Arrhenius prefactor $e^{4J/k_BT}$ grows dramatically for lower temperatures, reaching infinity with $T\to 0$~\cite{spirinFateZerotemperatureIsing2001}.
	The scaling with $L^3$ has also been verified with our data at $T=0.2T_c$ as illustrated in figure~\ref{fig:diagonal_stripes_tau_vs_L}.
	For sufficiently small $T>0$, the given percentages of each type of metastable state still apply approximately~\cite{spirinFreezingIsingFerromagnets2001}.

	This $L$-dependence of the stripes' lifetime is a central issue in this study.
	Since relation~(\ref{eq:L-scaling}) is fitted over all lattice sizes $L$, we must require that the stationary states are comparable between all of them.
	Ideally, this dictates full equilibration for all systems, as per condition (\ref{eq:C_inf_definition}).
	Whether and how this requirement can be loosened or even lifted will be explained more detailed in later subsections.
	This is also interesting for future work, as there are systems, especially any Ising lattice at dimension $d \geq 3$
	without external field or long-range interactions, which will almost never reach equilibrium after being quenched below $T_c$~\cite{agrawalAsymptoticStatesIsing2022,spirinFreezingIsingFerromagnets2001}.
	Consequently, it is important to assess here already, whether keeping the (horizontal or vertical) striped states in the data pool affects the estimations at all.
	Later on, several distinct possibilities of dealing with striped states will be explored as well.

\begin{figure}[t]
		\centering
		\includegraphics[width=0.7\linewidth]{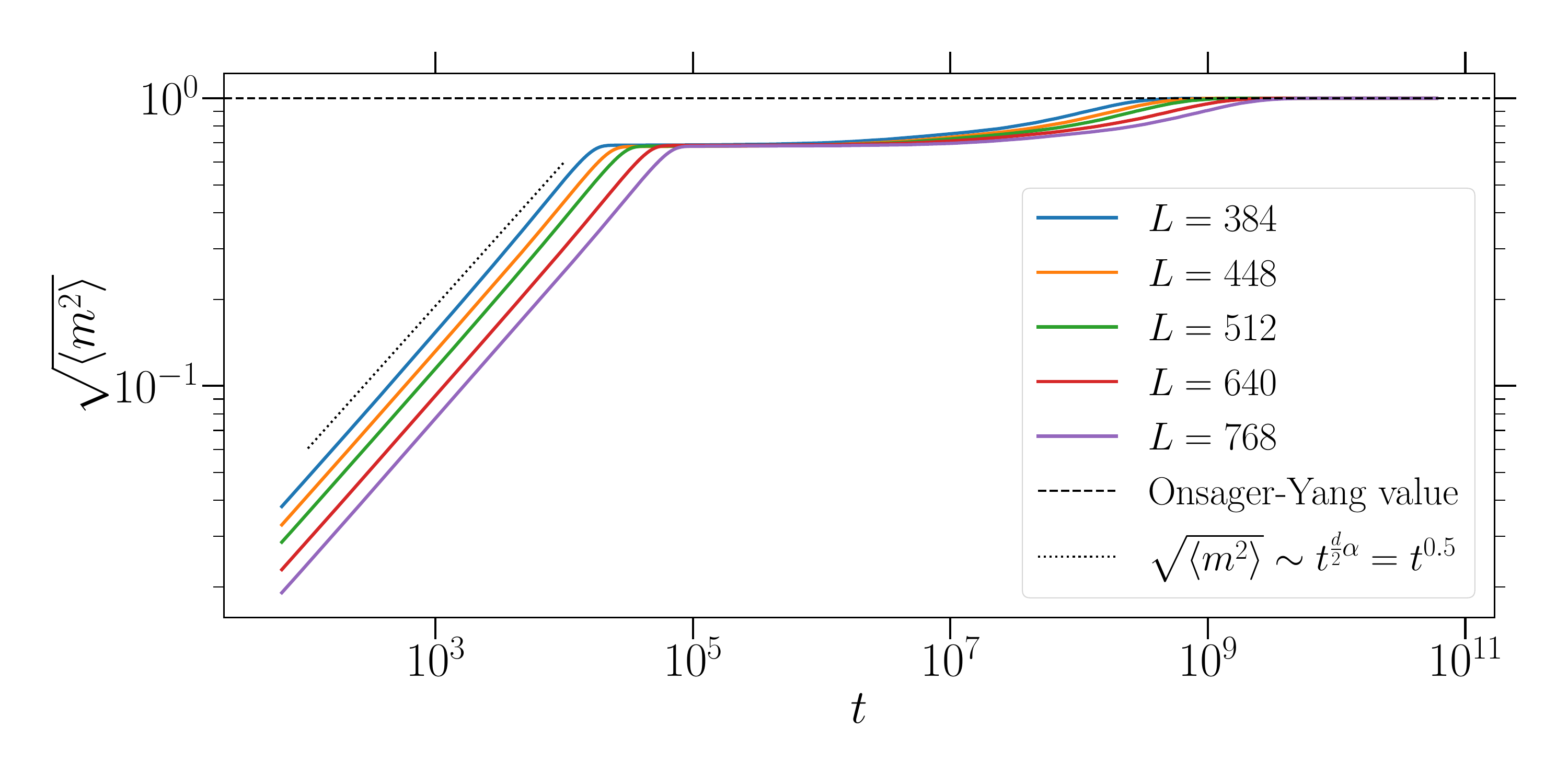}
		\caption{Square root of the average square magnetization for lattice sizes $L \in \{384, 448, 512, 640, 768\}$ at $T=0.2T_c$. During equilibration, the magnetization grows algebraically~\cite{jankeRoleMagnetizationPhaseordering2023} with $\sqrt{\langle m^2 \rangle} \sim t^{\frac{d}{2}\alpha}$ (where $\alpha = 1/z = 0.5$) as indicated by the dotted black line until it stalls at the pseudo-plateau owing to the longevity of metastable states. As time passes, the systems reach equilibrium, settling at the expected theoretical value given by the Onsager-Yang solution~\cite{yangSpontaneousMagnetizationTwodimensional1952,baxterOnsagerKaufmansCalculation2011}. Error bars have been omitted for clarity.}
		\label{fig:mag_0.2Tc}
	\end{figure}

	The consequences of these metastable stripe configurations on the plateau height $\plath$ are the reason for the step-wise relaxation seen in figure~\ref{fig:c-tsl768_0,2Tc_full_equilibration}.
	For moderately large times, the plateaus' heights seem to converge towards a constant value.
	However, the existence of metastable states causes the autocorrelator to exhibit clearly visible steps.
	Their existence is related to the size-dependent lifetime, see figure~\ref{fig:diagonal_stripes_tau_vs_L}, of all types of metastable states and the steps then occur as soon as the evaporation process begins.
	This cautions against a too simplistic approach and this effect might also serve as a rough estimate of the required simulation time scales.
	%============================================================================================================================%%
	%============================================================================================================================%%
	Following the dynamics for longer time then leads, above $T=0$, to \emph{all} stripes' evaporation.
	This gives the second and final plateau which produces the values of $\plath$ for our analysis.
	At this point, the system has fully equilibrated, which we have checked by comparing the magnetization $m$ to the expected value $m(\beta)=\left(1-\sinh^{-4}(2\beta)\right)^{1/8}$ found in the Onsager-Yang solution of the $2D$ Ising model~\cite{yangSpontaneousMagnetizationTwodimensional1952,baxterOnsagerKaufmansCalculation2011}, illustrated in figure~\ref{fig:mag_0.2Tc}.
	Consequently, we can be certain that condition~(\ref{eq:C_inf_definition}) is sufficiently fulfilled.

	%============================================================================================================================%%
	%============================================================================================================================%%
	\medskip

\begin{figure}[t]
		\centering
		\includegraphics[width=1\linewidth]{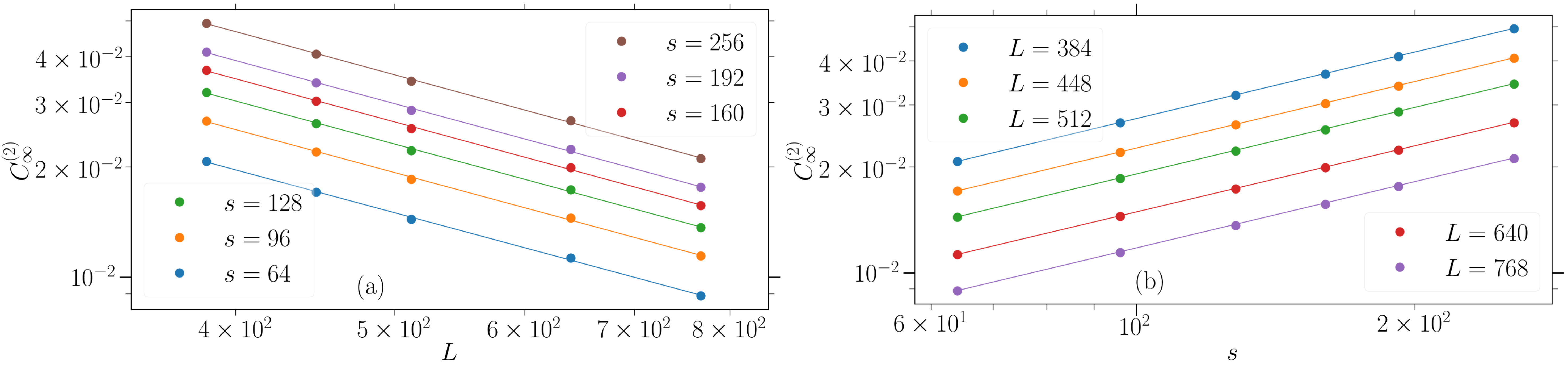}
		\caption{Plateau heights $\plath$ at $T=0.2T_c$ plotted against (a) the lattice size $L \in \{384, 448, 512, 640, 768\}$ with the solid line representing the fit using equation~(\ref{eq:L-scaling}); and (b) against the waiting time $s \in \{64, 96, 128, 160, 192, 256\}$ with the solid line representing the fit using equation~(\ref{eq:s-scaling}). The error bars are smaller than the symbol size.}
		\label{fig:withfit_0.2Tc}
	\end{figure}

	After having obtained the definitive plateau heights from equilibrium, the application of equations~(\ref{eq:L-scaling}) and (\ref{eq:s-scaling}) should be straightforward, by plotting $\plath$ versus either $L$ or $s$ as shown in figure~\ref{fig:withfit_0.2Tc}.
	%
	%============================================================================================================================%%
	%============================================================================================================================%%
	%
	In figure~\ref{fig:withfit_0.2Tc}(a), the variation of the plateau height $\plath$ with $L$ for several values of $s$ (extracted from figure~\ref{fig:c-tsl768_0,2Tc_full_equilibration}(a) by taking the final data point of each plateau) is shown.
	For each curve at a fixed $s$, an error-weighted fit of $C_\infty^{(2)} \sim L^{-\lambda}$ (cf. equation~(\ref{eq:L-scaling})) is performed.
	Thus, each $s$ provides in itself a valid estimate of $\lambda$ as the correlation between the waiting times $s$ is orthogonal to the fitting direction.
	Remarkably, all fits produce roughly parallel lines in this double logarithmic plot, such that we can be confident that the observed scaling behavior matches our predictions.
	To average over all single estimates of $\lambda$, however, the correlation in $s$ needs to be taken into account, such that we acquire a final estimate for $\lambda$ by combining the results for each fixed $s$ using the correlation-weighted average, see reference~\cite{weigelErrorEstimationReduction2010}.
	The numerical result is given by equation~(\ref{eq:estimate_full_equilibration}).

	Figure~\ref{fig:withfit_0.2Tc}(b) shows the variation of $\plath$ with $s$, for fixed values of $L$ (extracted from the final data points in figure~\ref{fig:c-tsl768_0,2Tc_full_equilibration}(b)), and is compared to the expected scaling $C_\infty^{(2)} \sim s^{\lambda/z}$ (cf. equation~(\ref{eq:s-scaling})).
	Here, the fitting procedure for each lattice size $L$ is more complex, as the correlation in $s$ spans right across the abscissa, such that the individual data points follow the relation closer than expected, leading to possibly inaccurate error estimates.
	As explained in section~\ref{sec:data_analysis}, this can be fully compensated by issuing a \emph{correlated} fit of $\plath$ vs $s$ for each fixed $L$ such that there is a proper estimate of $\lambda/z$ for each lattice size.
	These can then be combined into a single estimate using an error-weighted average, whose result is shown in equation~(\ref{eq:estimate_full_equilibration}) as well.

	\medskip

	Despite the ease of this method, whose complexity is given solely through the rectifiable temporal correlations in our data, there has to be awareness of \emph{which} specific range of $s$ and $L$ can be chosen, such that our central relations (\ref{eq:L-scaling}) and (\ref{eq:s-scaling}) still hold true.
	Thus, in order to mitigate systematic errors, several aspects have to be taken into account.

\begin{figure}[t]
		\centering
		\includegraphics[width=0.6\linewidth]{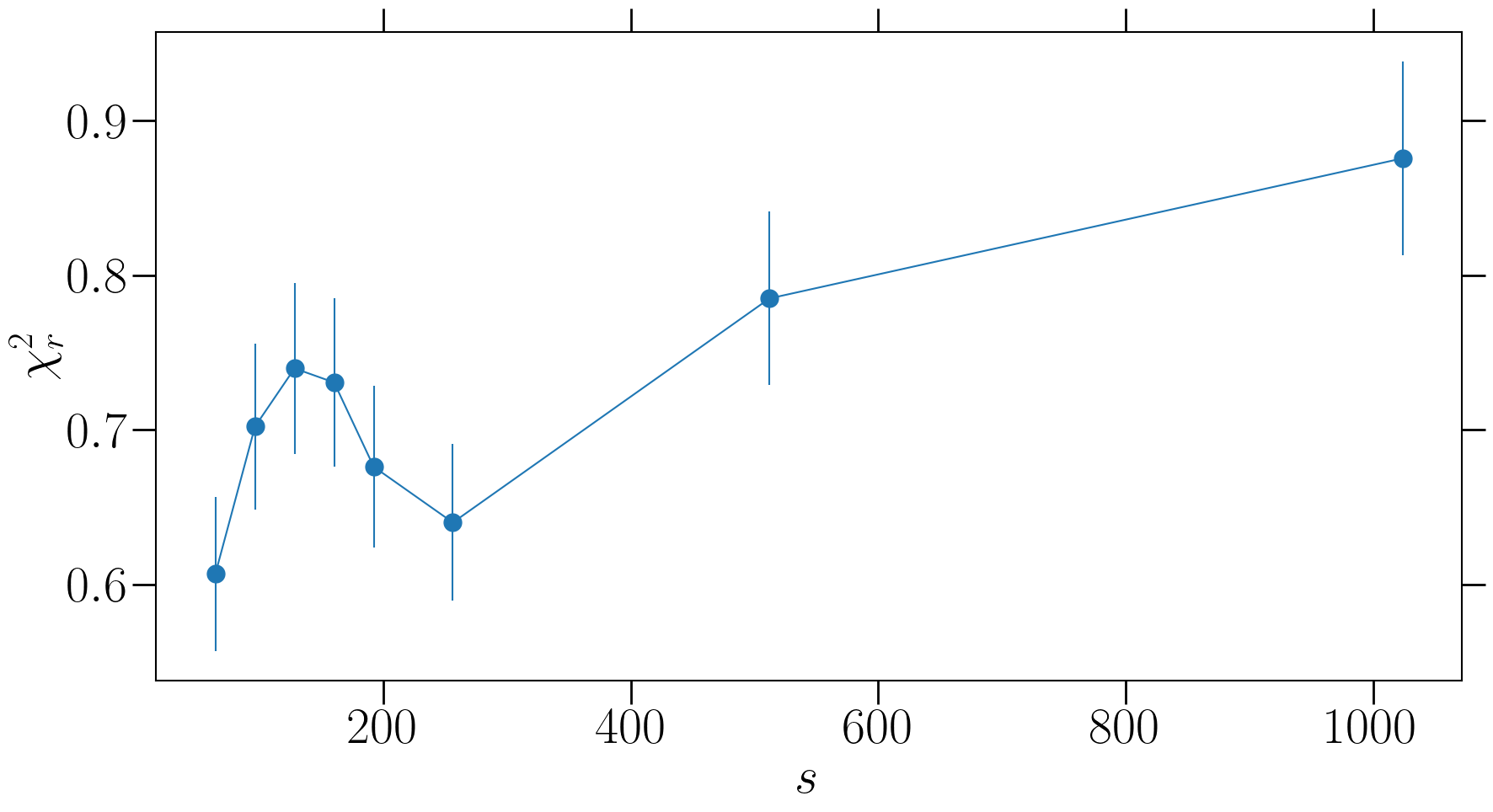}
		\caption{Reduced chi squared $\chi^2_r$ for the error-weighted fits of $\plath$ against $L$ to obtain $\lambda$ (similar to figure~\ref{fig:withfit_0.2Tc}(a)) using $L \in \{384, \allowbreak 448, \dots, \allowbreak 768\}$ and $s \in \{64, \allowbreak 96, \dots, 1024\}$. The increase of $\chi^2_r$ beginning at $s=512$ indicates the end of validity of the asymptotic formula (\ref{eq:s-scaling}) when including larger waiting times for the selected lattice sizes $L$.}
		\label{fig:chisquared}
	\end{figure}

	Some conditions to be imposed are straightforward.
	On one hand, since aging is a late-time phenomenon \cite{brayTheoryPhaseorderingKinetics2002,tauberCriticalDynamics2014,henkelAgeingDynamicalScaling2010,puriKineticsPhaseTransitions2009,mazenkoNonequilibriumStatisticalMechanics2008,cugliandoloDynamicsGlassySystems2002}, the waiting time $s$ must be chosen large enough.
	On the other hand, the waiting time should still be small enough that finite-size effects are not yet visible.
	Both conditions are met by choosing parameters in such a way that phenomenologically, the characteristic length scale is in the power-law scaling regime, i.e., $1 \ll \ell(s)\sim s^{1/z} \ll L$.
	Since deviations from this are not always clearly visible, one might try to detect them by observing the reduced chi squared $\chi^2_r = \chi^2/\mathrm{\#dof}$ with $\mathrm{\#dof}$ being the number of degrees of freedom at regression.
	This is done in figure~\ref{fig:chisquared} which uses (\ref{eq:s-scaling}) in a fit to our data set, with the inclusion of two additional waiting times $s=512$ and $s=1024$ for demonstrative purposes.
	For $s$ small enough, $\chi^2_r$ is roughly constant and well within an acceptable range, whereas a worsening tendency can be seen for the larger $s=512$ and $s=1024$ in combination with the given lattice sizes.
	This should mean that for smaller $s$ (or equivalently $L$ large enough), the assumed power-law fit is admissible, whereas for $s\gtrsim 512$ this may not be the case.
	Probably, finite-size effects become then noticeable in (\ref{eq:s-scaling}).
	In practice, both scaling relations must be checked, as these effects depend on the specific extremes of both, $L$ and $s$.

	%============================================================================================================================%%
	%============================================================================================================================%%

	\medskip

	Following the above guidelines, the parameter space $(s,L)$ for all subsequent results was fixed to the following sets:
	$L \in \{384,\allowbreak 448,\allowbreak 512,\allowbreak 640,\allowbreak 768\}$ and $s \in \{64, \allowbreak 96, \allowbreak 128, \allowbreak 160, \allowbreak 192, 256\}$ in order to minimize the systematic errors.

	\medskip
	
	%============================================================================================================================%%
	\begin{figure}[t]
		\centering
		\includegraphics[width=1\linewidth]{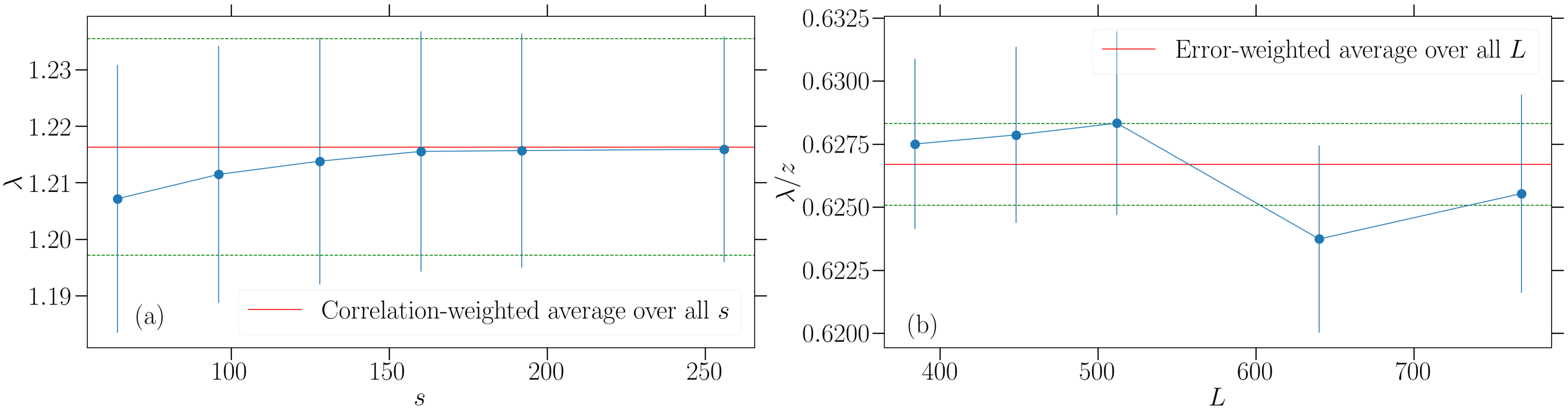}
		\caption{Estimates of (a) $\lambda$ and (b) $\lambda/z$ at $T=0.2T_c$ by fitting the curves of figure~\ref{fig:withfit_0.2Tc}
			and combining the estimates of the respectively fixed (a) $s$ using the correlation-weighted average and (b) $L$ using the error-weighted average.
			Owing to the intrinsic correlation of the chosen waiting times $s$, the correlation between the individual $\lambda$
			estimates is very high such that the combined estimate (red line, as given by equation~(\ref{eq:estimate_full_equilibration})) in (a) lies above all of the individual data points.
			The dashed green lines give the ($1 \sigma$) error estimate.}
		\label{fig:bothestimates0,2Tc_full_equilibration}
	\end{figure}
	%============================================================================================================================%%
	%
	After taking everything into account, the estimates resulting from each of our fits are displayed in figure~\ref{fig:bothestimates0,2Tc_full_equilibration}.
	Panel~\ref{fig:bothestimates0,2Tc_full_equilibration}(a) shows the individual $\lambda(s)$ for each fixed waiting time $s$, after having performed the error-weighted fit $C_\infty^{(2)} \sim L^{-\lambda}$ (cf. equation~(\ref{eq:L-scaling})), as blue dots with the respective $1\sigma$ error estimate per fit.
	To combine the individual estimates into a single value we use the correlation-weighted average which is shown as horizontal red line while its error estimate can be seen as dashed green lines above and below.
	Similarly, the individual $(\lambda/z)(L)$ are shown in panel~\ref{fig:bothestimates0,2Tc_full_equilibration}(b) for each lattice size $L$ resulting from correlated fits per $C_\infty^{(2)} \sim s^{\lambda/z}$ (cf. equation~(\ref{eq:s-scaling})).
	These estimates are combined into one by an error-weighted average.
	
	On the technical side, figure~\ref{fig:bothestimates0,2Tc_full_equilibration} shows several aspects worth highlighting in greater detail, which all point towards a common cause.
	We note that our estimates of $\lambda/z$ are significantly more precise than those of $\lambda$, which is true for both: the individual fits for each fixed $s$ or $L$ and the respective averages.
	In fact, the final estimate $\lambda$ has an at best marginally smaller error than its constituents, which is somewhat obvious at $s=256$ in panel~\ref{fig:bothestimates0,2Tc_full_equilibration}(a).
	This hints towards the fact that the amount of (new) statistical information per estimate $\lambda(s)$ may be rather small.
	Secondly, we can see a trend in panel~\ref{fig:bothestimates0,2Tc_full_equilibration}(a) where smaller values of $s$ produce smaller estimates of $\lambda$ until they saturate around $s \geq 160$.
	There, they coincide with the correlation-weighted average which, additionally, can be found beyond almost all individual data points, from which it is produced.
	This is particularly visible in comparison to the neighboring panel~\ref{fig:bothestimates0,2Tc_full_equilibration}(b) where none of these features occur.
	The individual estimates $(\lambda/z)(L)$ are randomly distributed around the average whose error estimate is smaller than the uncertainty of each data point.
	All of these counterintuitive issues can be traced back to the inherent correlations between the distinct waiting times $s_i$, see section~\ref{sec:data_analysis}, with a Pearson correlation coefficient of $\rho \approx 0.930 \dots 0.999$, depending on the distance in $s$.
	Especially the last aspect, i.e., the correlation-weighted average lying outside of the expected region, is a well-known phenomenon, often found in fits of correlated data~\cite{cannerCuriousResultsUsing1969}.
	To check whether these correlations could impact our results and their error estimates, we have repeated the same data analysis procedure while supplying each waiting time $s_i$ with a set of independent runs.
	More specifically, we divided our set of $30\,000$ runs into five equal non-overlapping parts, each of which is assigned to one waiting time.
	Then, each estimate for the plateau height $\plath(s_i,L)$ stems from $6\,000$ runs which are independent for each $s_i$ and, as a consequence, the resulting individual estimates $\lambda(s_i)$ can be combined via an error-weighted average as the correlations have been rectified by using these distinct data sets.
	We have seen that the aforementioned artifacts disappear completely while the results do not change significantly (except for the expected $\sqrt{5}$ increase of the error bars).
	Thus, we are confident that these effects do not falsify our estimates, in principle.
	We have also checked that the results do not change significantly by adding more waiting times in between as the correlations are high enough for additional $s$ inside the given range to not matter significantly.
	
	Since the correlations have been proven manageable, we take all runs into account for all $s$ throughout this work and do not use distinct data sets, i.e., we accept the intrinsic correlations as otherwise the number of simulations would scale with the number of waiting times.
	More specifically, we use our $30\,000$ runs and each run produces an estimate for every waiting time $s$ at once.
	As the correlation-weighted average and the correlated fit described in section~\ref{sec:data_analysis} provide mathematically sound remedies for the described issues, we can, thus, utilize our data to the fullest extent.

	After having discussed the individual data points, we read off the following numerical results for the averages of $\lambda$ and $\lambda/z$ in figure~\ref{fig:bothestimates0,2Tc_full_equilibration}:

	\begin{equation}
		\label{eq:estimate_full_equilibration}
		\frac{\lambda}{z} = 0.6267(17), \;\; \lambda = 1.216(20)
	\end{equation}
	where the standard error estimates indicate the $1 \sigma$ value as throughout this work.
	The resulting value $z=1.941(36)$ is not far off the expected $z=2$~\cite{brayGrowthLawsPhase1994} and clearly, our results are in agreement with the conjecture $\lambda \leq 5/4$, whereas the estimates by Liu and Mazenko~\cite{liuNonequilibriumAutocorrelationsPhaseordering1991} fall much further beyond our error bars.
	If we assert $z=2$, which is without greater doubt for the $2D$ Ising model, we can furthermore estimate
	\begin{equation}
		\label{eq:estimate_full_equilibration_z=2}
		\lambda^* = 1.2534(34)
	\end{equation}
	which readily supports the stronger hypothesis of $\lambda = 5/4$~\cite{fisherNonequilibriumDynamicsSpin1988}.
	
	Two comments on our results (\ref{eq:estimate_full_equilibration}) are in order: First, our method yields a \emph{direct} estimate of the autocorrelation exponent $\lambda$. This comes about since the relation~(\ref{eq:L-scaling}) only contains this exponent, independently of any value of $z$, and in contrast to (\ref{eq:dynamical_scaling}) which rather yields a value for $\lambda/z$. In the literature, only estimates of $\lambda/z$ or for $\lambda^*$ with an assumed value $z=2$ are quoted, with estimated errors of the same order as in our direct estimation of $\lambda$. Second, our estimate of $\lambda/z$ yields a significant reduction in statistical uncertainty compared to earlier works. This then leads to our improved estimate $\lambda^*$, see equation~(\ref{eq:estimate_full_equilibration_z=2}).

	To check the consistency of our direct estimates in equation~(\ref{eq:estimate_full_equilibration}) we perform a rescaling of the two-time autocorrelator using the determined exponent $\lambda/z$ against time $t$, see figure~\ref{fig:c-tsl768_0,2Tc_overlap-1}(a), which is a good {\it a posteriori} confirmation of our estimate.
	Remarkably, a good data collapse is seen for the majority of the plot and, most importantly, {\em both} plateaus.
	Similarly, figure~\ref{fig:c-tsl768_0,2Tc_overlap-1}(b) shows an alternative based on the scaling with $L$, as per (\ref{eq:L-scaling}), where both plateaus overlap as well, demonstrating the validity of our $\lambda$ estimate.
	To push things further, the autocorrelator can be rescaled utilizing both relations at once, see figure~\ref{fig:c-tsl768_0,2Tc_overlap-2}, such that in panel (a) all $L$ and all $s$ share the same two plateaus in the very same plot.
	What is missing to create a proper master curve, is the rescaling of time $t$ which can be derived from the behavior of the striped metastable states.
	There, two distinct time scales are relevant.
	At first, for smaller $t$, we look at the \emph{formation} of stripes scaling with $L^2$, as seen in figure~\ref{fig:c-tsl768_0,2Tc_overlap-2}(b).
	Combined with (\ref{eq:L-scaling}) and (\ref{eq:s-scaling}), this leads to an overlap at and around the beginning of the pseudo-plateau, close to $t/L^2 \approx 10^{-1}$.
	On the flip side, the descend from the pseudo-plateau to equilibrium near $t/L^2 \approx 10^3$ remains frayed.
	The second time scale, which becomes relevant at this very point, is governed by the \emph{evaporation} of the metastable states.
	Consequently, driving the point home, we produce, in panel (c), the final master curve of the two-time autocorrelator by rescaling the $t$-axis by $1/L^3$, which is the inverse average lifetime of all striped states such that not only the curves for all $s$ fall on top of each other but also for all $L$ along the entire late-time region.
	Somewhat surprisingly, this should mean that, in practice, it might not be too essential which plateau,
	see again figure~\ref{fig:c-tsl768_0,2Tc_full_equilibration}, is used for the analysis.
	One should keep in mind, though, that systematic errors may arise since the overlap around the pseudo-plateau may not be sufficient at a large enough time window.
	More specifically, figure~\ref{fig:c-tsl768_0,2Tc_overlap-1}(a) shows a generous overlap between $t \approx 10^5 \ldots 10^7$, indicating that scaling relation~(\ref{eq:s-scaling}) may work well within this region even without full equilibration.
	Contrarily, panel~\ref{fig:c-tsl768_0,2Tc_overlap-1}(b) shows a good overlap around the same region but upon closer inspection, it is visible how soon single curves deviate from the pseudo-plateau such that relation~(\ref{eq:L-scaling}) may not be observed equally well at all $t$.
	Figure~\ref{fig:c-tsl768_0,2Tc_overlap-2}(a) combines these cautious remarks succinctly: One should be able to utilize the plateau height of the autocorrelator at any time $t$ where an overlap is found.
	This is obviously and demonstrably true for the equilibrium value.
	To test this hypothesis for the pseudo-plateau it was necessary to simulate at a lower temperature in order to stall the equilibration of stripes long enough such that we could pick a proper point in time where sufficient overlap should be expected. 

	%============================================================================================================================%%
	\begin{figure}[t]
		\centering
		\includegraphics[width=1\linewidth]{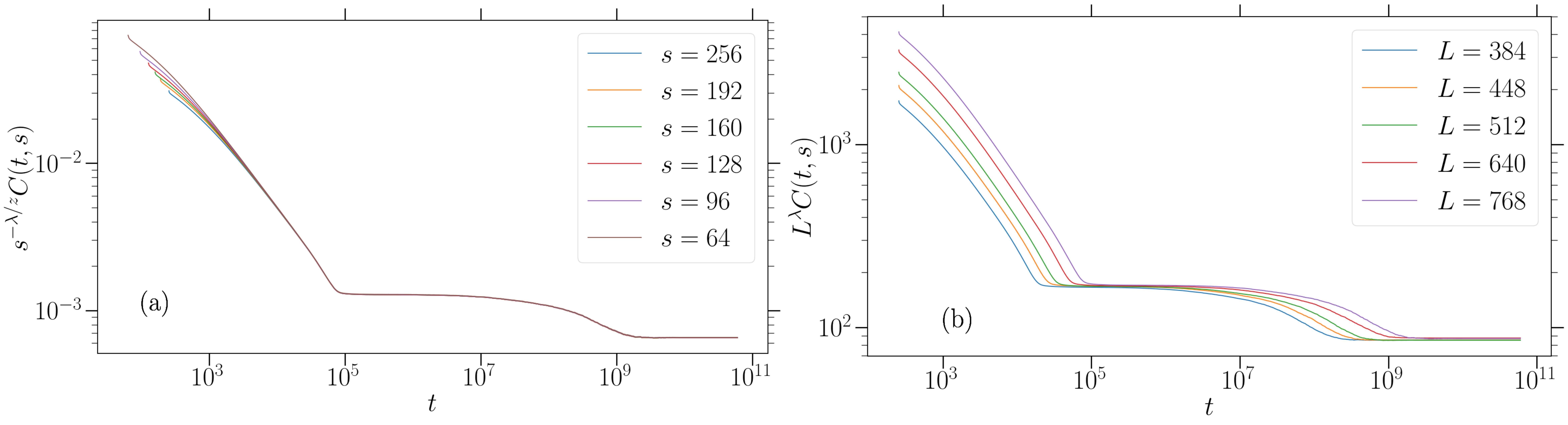}
		\caption{Rescaled two-time autocorrelator (a) $s^{-\lambda/z}C(t,s)$ at $L=768$, (b) $L^\lambda C(t,s)$ at $s=256$, both plotted versus time $t$ with the estimated exponents $\lambda/z \approx 0.6267(17)$ and $\lambda \approx 1.216(20)$. For clarity, error bars have been omitted.}
		\label{fig:c-tsl768_0,2Tc_overlap-1}
	\end{figure}
	%============================================================================================================================%%

	%============================================================================================================================%%
	\begin{figure}[t]
		\centering
		\includegraphics[width=1\linewidth]{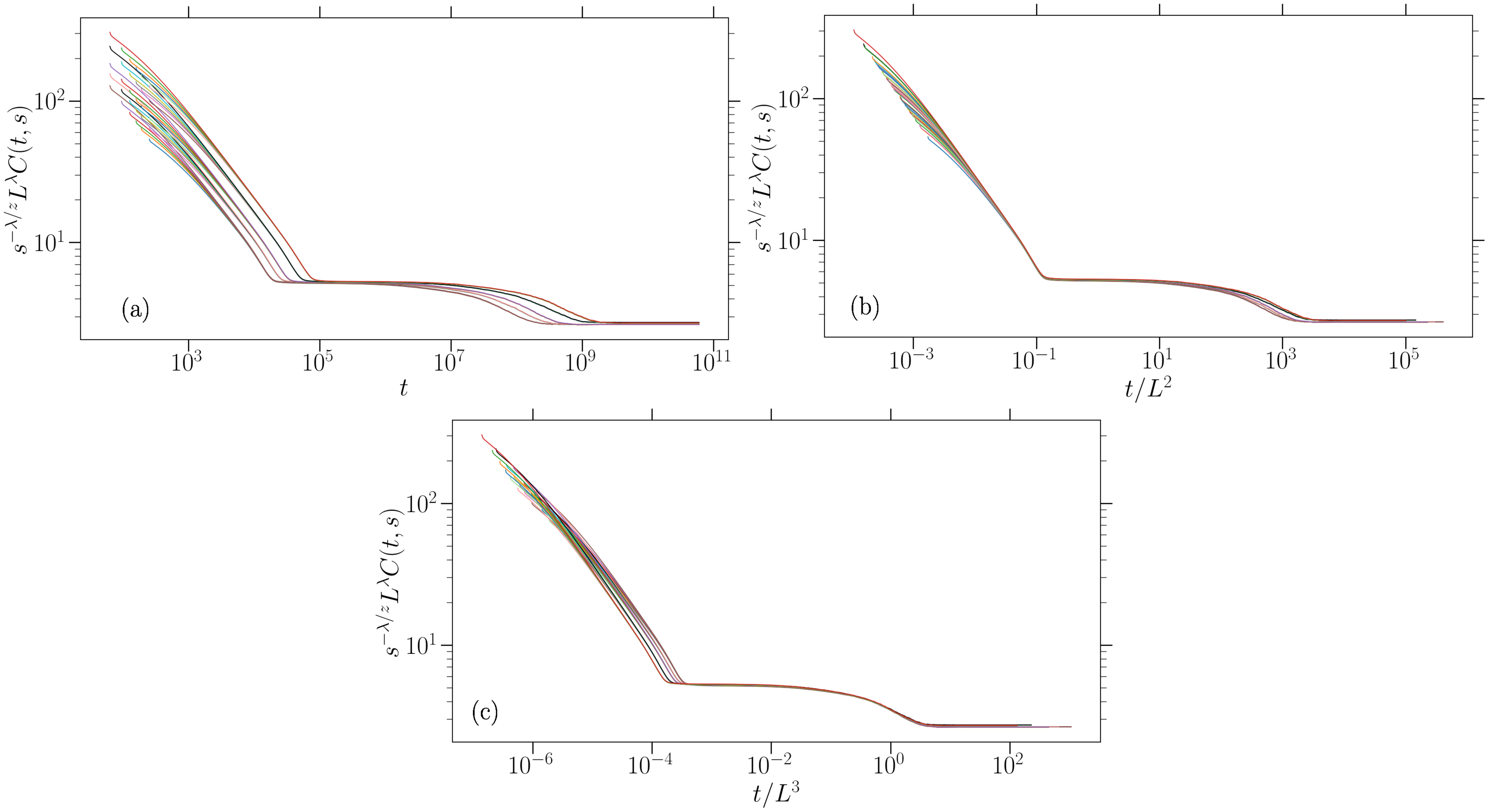}
		\caption{Rescaled two-time autocorrelator $s^{-\lambda/z}L^\lambda C(t,s)$ for all $L$ and $s$ at once plotted versus (a) time $t$ itself, (b) $t/L^2$ to scale with the time scale of stripe \emph{formation}, (c) $t/L^3$ to scale with the time scale of stripe \emph{evaporation}. For clarity, error bars have been omitted.}
		\label{fig:c-tsl768_0,2Tc_overlap-2}
	\end{figure}
	%============================================================================================================================%%

	\subsection{\label{subsec:0.1Tc} Estimates at \texorpdfstring{$T=0.1T_c$}{T=0.1Tc}}
	\subsubsection{\label{subsubsec:alldata}Including metastable states}

	%%============================================================================================================================%%
	\begin{figure}[t]
		\centering
		\includegraphics[width=1\linewidth]{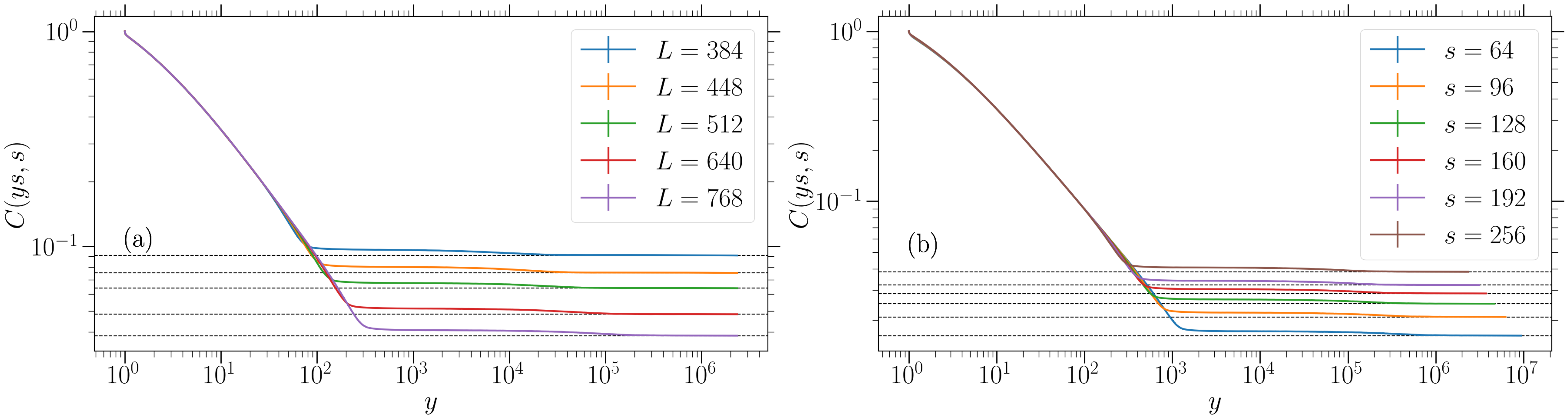}
		\caption{Two-time autocorrelator $C(ys,s)$ for the $2D$ Ising model quenched to $T=0.1T_c$ for various fixed (a) lattice sizes $L$ at $s=256$, or (b) waiting times $s$ at $L=768$.
			In comparison to figure~\ref{fig:c-tsl768_0,2Tc_full_equilibration}, the simulation has ended after $6 \times 10^8$ sweeps before equilibrium could be reached, such that \emph{neither} of the two plateaus visible represents the long-time behavior per equation~(\ref{eq:C_inf_definition}).
			Here, the first pseudo-plateau corresponds to an ensemble of states including all kinds of stripes (viz. figure~\ref{fig:stripes}). It starts at a similar point in time as for $T=0.2T_c$. After a small step at around $y \approx 10^4 \ldots 10^5$ (depending on $L^3$) in panel (a), all \emph{diagonal} stripes have evaporated, leaving the horizontal and vertical behind. This causes the second pseudo-plateau. }
		\label{fig:c-tsl768_0,1Tc}
	\end{figure}
	%%============================================================================================================================%%

	In figure~\ref{fig:c-tsl768_0,2Tc_overlap-2}, we observed two distinct plateaus for $T=0.2T_c$ which both suggest a noticeably similar scaling behavior.
	A similar although in its core far different situation can be observed for the two-time autocorrelator at $T=0.1T_c$.
	There, upon closer inspection, a pseudo-plateau is visible once again as seen in figure~\ref{fig:c-tsl768_0,1Tc} in the range of $y \approx 10^3 \ldots 10^4$.
	The dashed black lines indicate the plateau heights at the last data point from which the first pseudo-plateau is clearly offset.
	As the temperature is significantly lower, the metastable states' lifetime extends far beyond what was simulated in this case.
	Thus, what is visible here at the end of $C(ys,s)$ is not the \emph{actual} plateau as per equation~(\ref{eq:C_inf_definition}) as there are still the usual percentages of metastable states included, such that we are in a stationary state instead of equilibrium.
	More specifically, the behavior of $C(ys,s)$ in figure~\ref{fig:c-tsl768_0,1Tc} follows the same trajectory as with $T=0.2T_c$, until it hits the first pseudo-plateau around $y \approx 10^2$ (for $L=384$) in figure~\ref{fig:c-tsl768_0,1Tc}(a).
	There, all kinds of metastable states are part of the ensemble such that the autocorrelator stalls until these long-living states vanish one after another.
	This process can be observed by following the graph until around $y \approx 10^4 \ldots 10^5$ (depending on $L^3$) where the autocorrelation exhibits a noticeable step downwards, which is the point where the majority of diagonal stripes has evaporated, until none are left at the end of the second pseudo-plateau visible.
	It should be noted that a similarly strong distinction between these two pseudo-plateaus is not possible at $T=0.2T_c$, since the different lifetimes of diagonal and straight stripes are not as far off from each other as necessary, owing to the strong $T$-dependence of the Arrhenius prefactor $e^{4J/k_BT}$ in the $\tau_\mathrm{Stripe}$ scaling relation.
	Thus, what is distinctly visible here as two plateaus is not discernible in figure~\ref{fig:c-tsl768_0,2Tc_full_equilibration} as there, the first pseudo-plateau merges directly into the final one while diagonal and straight stripes evaporate on a similar time scale.

	In order to test if it is legitimate to	concentrate on the second pseudo-plateau in figure~\ref{fig:c-tsl768_0,1Tc}, where only the diagonal stripes evaporated while horizontal and vertical stripes still remain, we now analyze what happens after a quench to the lower temperature $T=0.1 T_c$, with the simulation running for $t=6\times10^{8}$ sweeps,
	such that all diagonal stripes have vanished while the portion of horizontal and vertical stripes remains.
	This temperature has been chosen specifically, as it produces a pseudo-plateau large enough to comfortably pick a point in time where the number of striped states mirrors the percentages mentioned before in section~\ref{subsec:0.2Tc} with sufficient accuracy for all $L$ and $s$.

	Taking the final point in $C(ys,s)$ in figure~\ref{fig:c-tsl768_0,1Tc} as `plateau height' and utilizing the same analysis routine as before, we then find
	\begin{equation}
		\label{eq:estimate_all_stripes}
		\frac{\lambda}{z} = 0.6212(9), \;\; \lambda = 1.226(11)
	\end{equation}
	and consequently $z=1.974(20)$ and $\lambda^* = 1.2423(18)$ (assuming $z=2$).

	A comparison of this with the earlier estimates~(\ref{eq:estimate_full_equilibration}) for $T=0.2T_c$ shows statistically comparable results, and again is in agreement with \cite{fisherNonequilibriumDynamicsSpin1988}.
	Moreover, the lower temperature produces only half the statistical error with the same number of runs, such that the thermal noise affecting the plateau heights can be identified as the primary source of error.
	Depending on the precise circumstances and algorithmic parameters, the required wall-clock time at lower $T$ can be comparable to higher $T$, as the increase in the number of necessary equilibration sweeps, caused by the temperature-dependence of $\tau_\mathrm{Stripe}$, may be almost fully compensated by an equally increased algorithmic performance, owing to the rising efficiency of the $n$-fold way algorithm at lower $T$, as we have seen in our simulations.

	\subsubsection{\label{subsubsec:nostripes}Removing metastable states}

	Since about $34\%$ of all runs, after the quench, end up in a metastable state without diagonal stripes and are weakly dependent on the lattice size $L$, we now inquire if and how all striped configurations can be purged. {\it A priori}, the effect on the determinations of $\lambda/z$ and $\lambda$ is not entirely clear,
	although figures~\ref{fig:c-tsl768_0,2Tc_full_equilibration} and~\ref{fig:c-tsl768_0,2Tc_overlap-2} show that all plateaus scale in a similar manner.

	Indeed, estimates only from the $\approx 62\,\%$ fully aligned configurations are statistically compatible with the former ones and we find
	\begin{equation}
		\frac{\lambda}{z} = 0.6259(15), \;\; \lambda = 1.242(16)
	\end{equation}
	which would imply $z=1.984(30)$ and $\lambda^* = 1.2519(29)$ (assuming $z=2$).

	Comparing with our estimates (\ref{eq:estimate_full_equilibration}) at $T=0.2T_c$ as our primary result without special techniques shows that overall, the results have not changed significantly and still agree with the conjecture $\lambda = 5/4$ \cite{fisherNonequilibriumDynamicsSpin1988}.
	Still, there is a noticeable upwards shift for the $\lambda$ estimate.

\begin{figure}[t]
		\centering
		\includegraphics[width=0.7\linewidth]{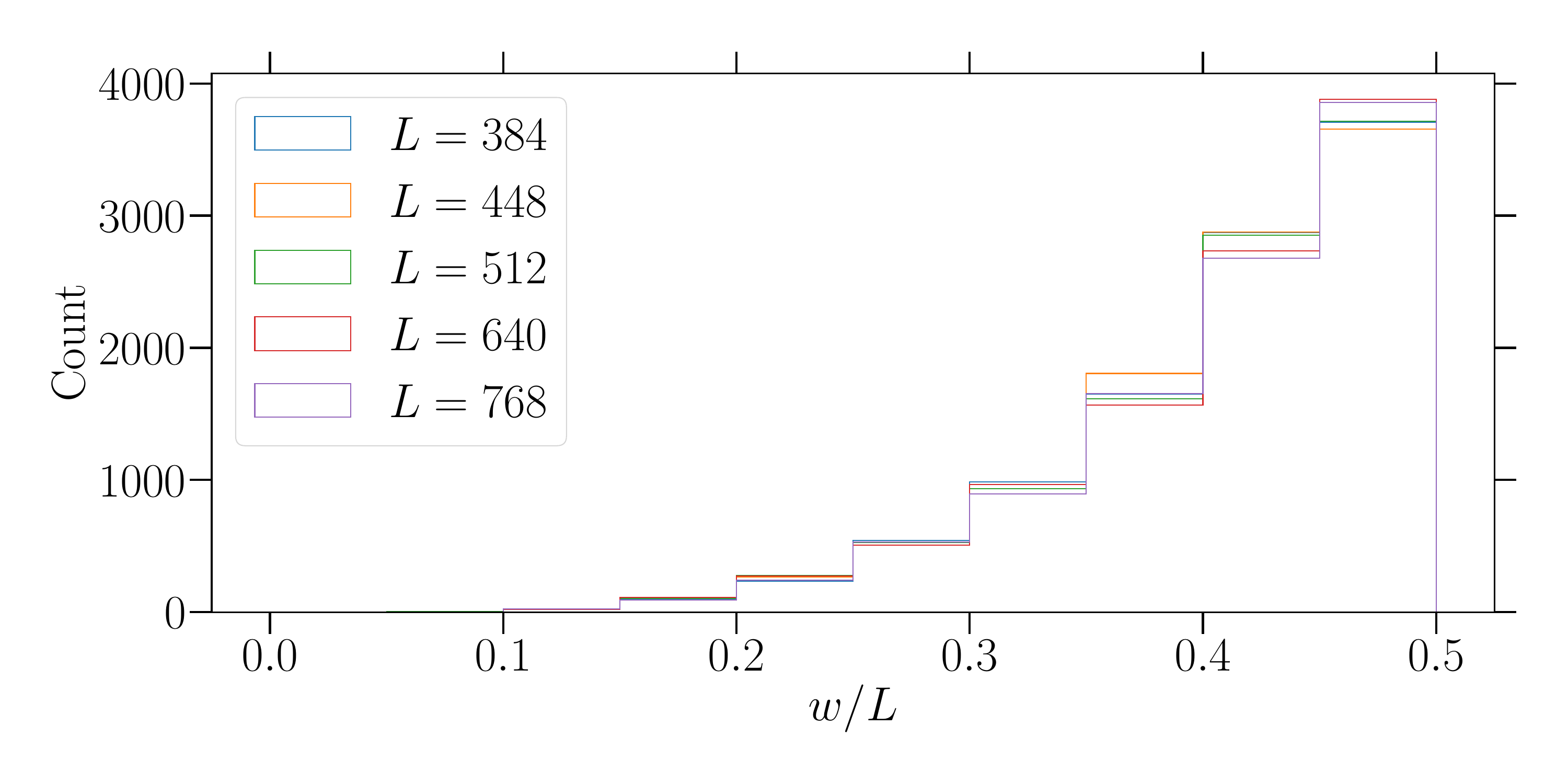}
		\caption{Frequency of relative stripe width ${w(L)}/{L}$ for all simulated lattice sizes at $T=0.1T_c$.}
		\label{fig:stripe_width_histogram}
	\end{figure}

	Conceptually, this agreement with the results in (\ref{eq:estimate_all_stripes}) can be understood as follows.
	If the domains' growth is limited by the system size, it should be equally limited by another domain.
	Then the finite-size scaling formula~(\ref{eq:both-scalings}) is extended to
	\begin{equation}
		\label{eq:total-scaling_with_stripes}
		\plath \sim \left(\frac{L-w(L)}{s^{1/z}}\right)^{-\lambda}
	\end{equation}
	where $w(L)$ is the width of a (perfectly straight) stripe of the minority spin value, depending on system size $L$.
	The $s$-scaling of $\plath$ is obviously unaffected, assuming that $w(L) = c L \sim L$.
	Comparing equation~(\ref{eq:total-scaling_with_stripes}) with the expected $L$-scaling (\ref{eq:L-scaling}) of $C^{(2)}_{\infty}$ yields
	\begin{equation}
		{\plath}  \sim L^{-\lambda}  \left(1 - \frac{w(L)}{L}\right)
	\end{equation}
	which deviates from equation~(\ref{eq:L-scaling}) with respect to the \emph{relative} stripe width.
	A histogram of $w/L$ is shown in figure~\ref{fig:stripe_width_histogram}, for all utilized $L$.
	Clearly, the system size has almost no influence on its shape.
	Therefore, we can conclude that our assumption $w(L) = c L$ is admissible, with $c=\mathrm{const.}$ for all $L$, such that $L-w(L) = (1-c)L$ leading, with equation~(\ref{eq:total-scaling_with_stripes}), to $\plath \sim L^{-\lambda}$, just as before.
	Thus, the distribution of the width $w(L)$ should not be important in the scaling of $C(ys,s)$
	and the stripes should not influence the values of $\lambda/z$ and $\lambda$.

	%============================================================================================================================%%
	%============================================================================================================================%%

	However, the above conclusion depends crucially on the assumption that any emerging striped states are just as straight as the lattice itself.
	Only if this is the case, it is legitimate to compare the finite-size effects coming from the domain growth up to its very borders to the finite-size effects of the domain growth up to another major domain wall.
	This assumption clearly fails as soon as any diagonal stripes emerge where relation~(\ref{eq:total-scaling_with_stripes}) cannot be applied,
	such that the evaporation (or transformation into horizontal or vertical stripes) of these states, which can always be expected at all $T$,
	draws a clear requirement for the minimum simulation time.
	In any case, having shown that horizontal and vertical stripes do not gravely influence our estimations, we can conclude that fully simulating until the equilibrium state is reached may not be, contrarily to equation~(\ref{eq:C_inf_definition}), a firm requirement.
	Thus, we can conclude that straight domain walls cause \emph{premature} finite-size effects comparable to those owing to finite lattice size.
	Whether this principle also extends to higher dimensions, especially the $3D$ Ising model is not as clear, though, as there rarely are comparably clear interfaces.

	%++++++++++++++++++++++++++++++++++++++++++++++++++++++++++++++++++++++++++++++++++++++++++++++++++++++++++++++++++++++++++++%%

	\subsubsection{\label{subsubsec:equilibratingstripes}Stochastically equilibrated striped states}

	A $2D$ Ising system is usually dominated by the majority phase right after the quench~\cite{barrosFreezingStripeStates2009}.
	Hence one may predict the equilibration outcome at low temperature given its occupation ratio on the lattice.
	Applied to our technique, this would allow for much shorter simulation times than waiting for all stripes to evaporate while conserving the necessary homogeneity over all $L$.

	%============================================================================================================================%%
	\begin{figure}[t]
		\centering
		\includegraphics[width=0.7\linewidth]{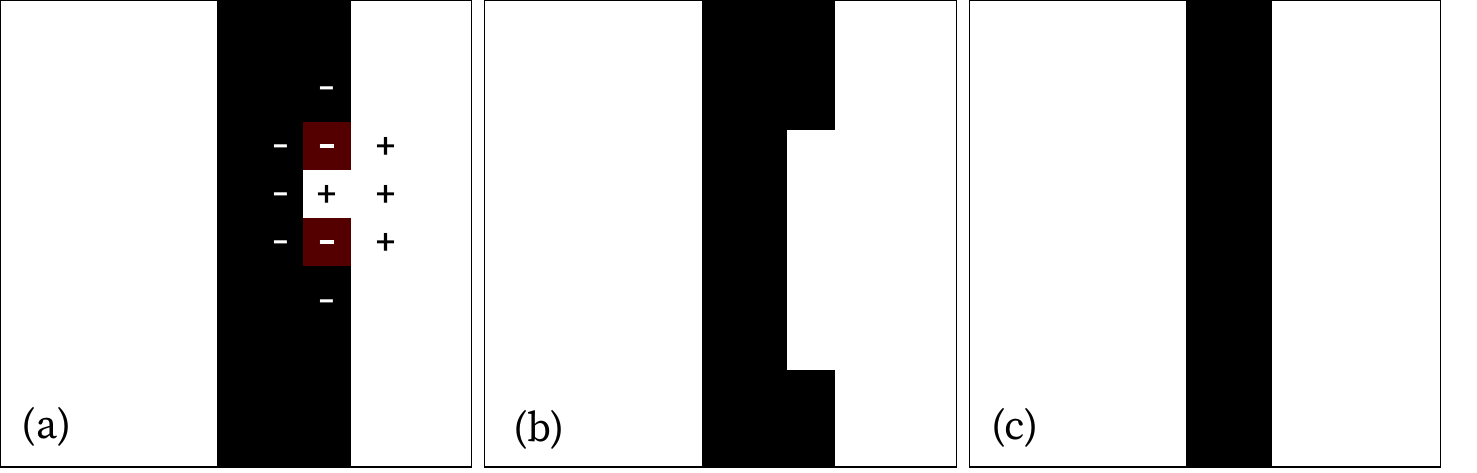}
		\caption{Step-wise evaporation of a stripe at $T>0$. This process also applies with roles reversed, i.e., the growth of a stripe by one column, such that the chance of a spin value dominating the lattice only depends on the respective stripe's width.}
		\label{fig:stripe_evaporation}
	\end{figure}
	%============================================================================================================================%%

	Based on the work by Spirin {\it et al.}~\cite{spirinFreezingIsingFerromagnets2001},
	the disappearance of a straight stripe at small $T>0$ can be described in several steps as illustrated in figure~\ref{fig:stripe_evaporation}.
	First, the stripe's outermost column of spins is interrupted by a single dent, thus creating a spin with three unlike neighbors such that the time necessary for this to happen is of the order of $e^{{4J}/{k_BT}}$.
	This dent then frees its two neighbors on the same layer which can now flip without energy cost, as both have two like and two unlike neighboring spins,
	viz. figure~\ref{fig:stripe_evaporation}(a).
	As a consequence, this broken up layer can now be described as a $1D$ random walk with $p=0.5$
	for a move in either direction, which terminates as soon as the interfaces meet,
	flipping either the whole layer or returning to the original state, i.e., exhibiting an absorbing boundary condition.
	The probability for a single layer to flip is ${1}/{L}$ according to first-passage properties, such that $L$
	dent creation events are necessary to change the stripe's width by $\pm 1$.
	Most interestingly, this description remains valid independently of the spin's value and whether the stripe shrinks or grows.
	Thus, as long as all stripes are straight, the probability for a metastable state to align with the spin value of a stripe of width $w$
	is simply\,\footnote{The probability is not exactly equal as a stripe of width $w=1$ almost always evaporates eventually after suffering a dent, i.e., a break.}
	$p_\mathrm{Stripe} \approx {w}/{L}$.
	Numerically, this can be implemented by measuring the minority phase's width $w$, then picking a random number $r \in [0,1)$
	and returning a fully aligned lattice with the minority's spin value, if $r < {w}/{L}$.

	%============================================================================================================================%%
	\begin{figure}[t]
		\centering
		\includegraphics[width=0.7\linewidth]{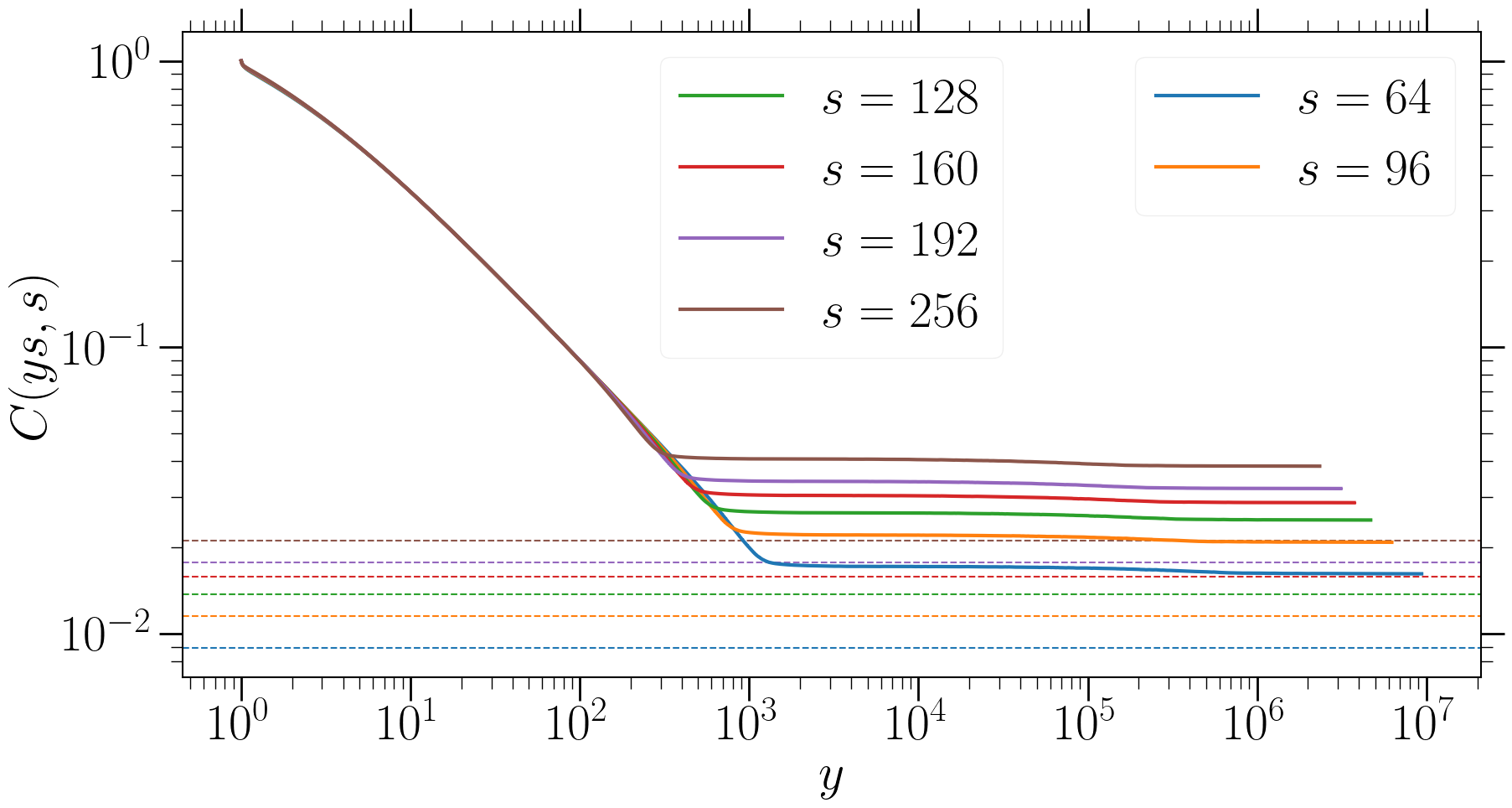}
		\caption{Two-time autocorrelator for $L=768$ at $T=0.1T_c$,
			where all horizontal or vertical stripes have been stochastically equilibrated according to the probability $p_\mathrm{Stripe} \approx {w}/{L}$.
			The dashed lines represent the final plateau height $C^{(2)}_{\infty}$ after equilibration for the respective waiting time $s$.}
		\label{fig:c-tsl768_equilibrated_stripes}
	\end{figure}
	%============================================================================================================================%%

	The impact of artificially equilibrating all stripes of the data at $0.1T_c$ can be observed in figure~\ref{fig:c-tsl768_equilibrated_stripes},
	where the final values for $C^{(2)}_{\infty}$ (given by the colored dashed lines) now lie far below	their respective plateaus, similarly to the results at $T=0.2T_c$.
	In fact, comparing figures~\ref{fig:c-tsl768_0,2Tc_full_equilibration} and \ref{fig:c-tsl768_equilibrated_stripes}, we can see that the plateau heights are almost equal for both temperatures, such that it is to be expected that the exponents' estimation will yield similar results.

\begin{figure}[t]
		\centering
		\includegraphics[width=1\linewidth]{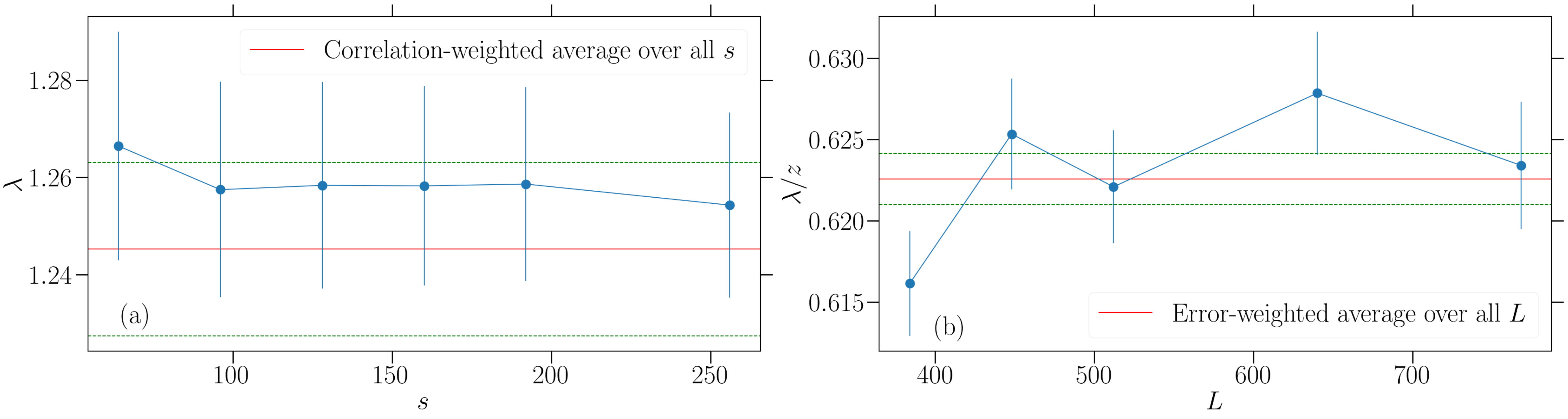}
		\caption{Estimates of (a) $\lambda$ and (b) $\lambda/z$ with all horizontal or vertical stripes stochastically equilibrated at $T=0.1T_c$.}
		\label{fig:bothestimatesequilibratedstripes}
	\end{figure}

	And indeed, carrying out the usual fits, we find
	\begin{equation}
		\label{eq:estimate_equilibrated_stripes}
		\frac{\lambda}{z} = 0.6226(16),  \;\; \lambda = 1.245(18)
	\end{equation}
	as visualized in figure~\ref{fig:bothestimatesequilibratedstripes} and thus $z=2.000(34)$ respectively $\lambda^* = 1.2452(32)$ (assuming $z=2$).
	%============================================================================================================================%%
	%============================================================================================================================%%
 	These estimates being close to (\ref{eq:estimate_full_equilibration}) clearly demonstrate the statistical equivalence of this method with all former estimates, thus presenting another viable solution to deal with metastable states.
	Nevertheless, it must be kept in mind that the set of configurations may only contain either fully aligned lattices or straight stripes,
	as otherwise the mechanism presented above does not hold any longer.
	It can be assumed~\cite{spirinFreezingIsingFerromagnets2001} that this trick remains applicable for $T < 0.2T_c$,
	as long as the simulation time is long enough to evaporate any diagonal stripes.
	At any higher temperature, the other techniques described above remain available.

	\subsection{\label{subsec:discussion}Discussion}

	The results of the preceding subsections have illustrated that our method may be well-suited to estimate $\lambda/z$ and $\lambda$ itself directly.
	As outlined in section~\ref{subsec:0.2Tc}, there are several sources of bias which have to be kept in mind, however, in order to reach the desired accuracy, even if the given formulae~(\ref{eq:L-scaling}) and (\ref{eq:s-scaling}) facilitate their proper gauge.
	Beside the necessity to assemble the specific ranges for waiting time $s$ and lattice size $L$
	where all underlying conditions hold, the most significant influence has been found to lie in metastable states as they introduce
	\emph{premature finite-size effects} disturbing the necessary homogeneity in all runs between all selected $L$.
	Their effects and how they influence our method are somewhat unexpected since our theoretical framework~\cite{henkelNonequilibriumRelaxationsAgeing2023} is to a great extent based on the spherical model, where these issues are of no concern.
	Unfortunately, the spherical model is special in this regard, as many other models in common use (especially outside of pure analytics) can exhibit similar, if not worse, degrees of metastability, such as the Ising model on the honeycomb lattice~\cite{takanoOrderingProcessKinetic1993,cheonOrderingKineticsIsing2004,blanchardCriticalPercolationDynamics2017}, the Potts model~\cite{ferreroLongtermOrderingKinetics2007} or the time-dependent Ginzburg-Landau equation~\cite{olejarzFate2DKinetic2012}.
	For the most simple case, the $2D$ Ising model we have presented several possible solutions and circumventions as numerical experiments whose estimates all deliver statistically comparable figures.
	Nevertheless, the case of very low temperature with a significant number of metastable states remains conceptually problematic, especially since the actual magnitude of their lifetime $\tau_\mathrm{Stripe}$ is rather obscure, such that the practical applicability of these attempts is not always universal, and in some systems, e.g., $3D$ Ising model, difficult to carry out.
	Thus, in most applications it should be the safest way to fully equilibrate the system under scrutiny, while the influence of temperature on the plateau heights' variance should be kept in mind.
	Owing to the correlations between the different waiting times which truly reduce the statistical uncertainty, the indirect measurement via $\lambda/z$, see equation~(\ref{eq:s-scaling})
	is preferable, instead of $\lambda$ directly via equation~(\ref{eq:L-scaling}).
	Consequently, to apply this method onto real-world experiments, the measurement of the two-time autocorrelator must simply be extended to reach the full plateau,
	using system sizes which are not near equilibration at the given waiting times.
	Then, after measuring the plateau heights, equation~(\ref{eq:s-scaling}) should give $\lambda/z$.
	Although the uncertainty given by a direct estimation of $\lambda$ is significantly higher, the point that this method allows an immediate access to the autocorrelation exponent still stands, and remains especially valid for systems, where $z$ is not as well-proven as for the Ising model shown here.
	If the system exhibits metastable states comparably smooth with respect to the $2D$ Ising model such that equation~(\ref{eq:total-scaling_with_stripes}) is expected to hold,
	then full equilibration can also be avoided with one of the techniques presented in this work while delivering comparable results, given the influence of metastable states on $\plath$ is comparable for all investigated $L$ at the chosen $t$.
	This requires a sufficiently low temperature (here around $0.1T_c$), however, as the pseudo-plateaus in $C(ys,s)$ quickly break down.

	%%%%%%%%%%%%%%%%%%%%%%%%%%%%%%%%%%%%%%%%%%%%%%%%%%%%%%%%%%%%%%%%%%%%%%%%%%%%%%%%%%%%%%%%%%%%%%%%%%%%%%%%%%%%%%%%%%%%%%%%%%%%%
	\section{\label{sec:conclusion}Conclusion and Outlook}
	%%%%%%%%%%%%%%%%%%%%%%%%%%%%%%%%%%%%%%%%%%%%%%%%%%%%%%%%%%%%%%%%%%%%%%%%%%%%%%%%%%%%%%%%%%%%%%%%%%%%%%%%%%%%%%%%%%%%%%%%%%%%%

	\begin{table}[t]
		\caption{Collection of all estimates for $\lambda$, $\lambda/z$ and the $z$ that follows from all temperatures and techniques used. The column $\lambda^*$ shows the estimate of the autocorrelation exponent using $\lambda/z$ with $z=2$ assumed.}
		\centering
		\begin{tabular}{c c l l l l}
			\hline
			$T$ & Metastable states & $\lambda/z$ & $\lambda$ & $z$ & $\lambda^*$ \\
			\hline
			$0.2T_c$ & None & $0.6267(17)$ & $1.216(20)$ & $1.941(36)$ & $1.2534(34)$ \\
			$0.1T_c$ & Included & $0.6212(9)$ & $1.226(11)$ & $1.974(20)$ & $1.2423(18)$ \\
			$0.1T_c$ & Excluded & $0.6259(15)$ & $1.242(16)$ & $1.984(30)$ & $1.2519(29)$ \\
			$0.1T_c$ & Stoch. equilibrated & $0.6226(16)$ & $1.245(18)$ & $2.000(34)$ & $1.2452(32)$ \\
			\hline
		\end{tabular}
		\label{tab:results}
	\end{table}

	We have presented a new approach to measure the autocorrelation exponent $\lambda$ and dynamical exponent $z$
	via a novel finite-size scaling relation for the plateau height $\plath$ in the large-$y$ limit of the two-time autocorrelator $C(ys,s)$
	in phase-ordering kinetics, after a quench to $0<T<T_c$.
	Its main point comes from the deliberate reliance on small lattice sizes, in contrast to conventional methods relying on the asymptotics of the scaling function
	$f_C(y)=C(ys,s)$, for $s$ large enough.
	We have detailed various aspects of our new approach and systematically analyzed, when and how it can be applicable for numerical and experimental use.
	It may be hoped that the speed-up achieved by limiting the number of spins to simulate should easily surpass any mere algorithmic improvements regarding conventional non-equilibrium simulations.
	Even though a considerable number of distinct initializations is necessary to reach a desirable precision in comparison, these usually profit from concurrency in a trivial manner.
	Furthermore, the mathematically simple structure of its central scaling relations (\ref{eq:L-scaling}),~(\ref{eq:s-scaling}) allows to identify and to correct multiple sources of bias which might be more difficult in direct fits of the scaling function $f_C(y)$.
	An important source of systematic errors stems from the possible occurrence of metastable states~\cite{barrosFreezingStripeStates2009,olejarzFate2DKinetic2012,spirinFateZerotemperatureIsing2001,spirinFreezingIsingFerromagnets2001,agrawalAsymptoticStatesIsing2022} and we discussed in detail the merits of several possibilities to deal with these.
	As we have discussed extensively, their behavior remains quite difficult to grasp, especially in other, more complex, systems, such that the optimal choice simply is to facilitate their evaporation by working with not too low a temperature after the quench, but still far enough away from the cross-over region near to the critical temperature $T_c$.
	In principle, the greater framework of this method should be easily transferable to other systems, where, conceptually, it may provide estimates independent of other techniques and can at the very least serve as an useful \emph{a posteriori} comparison.

	We have used the $2D$ Ising model for a non-trivial test for the computation of the autocorrelation exponent.
	All of our exponent estimates are collected in table~\ref{tab:results}.

	Our main result is given by its first line (see also equations~(\ref{eq:estimate_full_equilibration}) and (\ref{eq:estimate_full_equilibration_z=2})) stemming from the simulations at $T=0.2T_c$ where the \emph{true} plateau scaling without any auxiliary techniques was utilized.
	We have checked the internal compatibility of our data by considering the re-scaled autocorrelator, see figures~\ref{fig:c-tsl768_0,2Tc_overlap-1} and \ref{fig:c-tsl768_0,2Tc_overlap-2}.
	Furthermore, since the individual estimates $\lambda$ and $\lambda/z$ can be combined to obtain an estimate for the dynamical exponent $z$, its value can be utilized as another anchor for an \emph{a posteriori} check against known results, if available.
	For the kinetic Ising model of dimension $d \geq 2$, there exists the recently proven~\cite{masaokaRigorousLowerBound2025} rigorous lower bound $z \geq 2$, consistent with the well-know earlier observation of $z=2$~\cite{brayGrowthLawsPhase1994,brayTheoryPhaseorderingKinetics2002}.
	As table~\ref{tab:results} shows, this bound is found well within at most $1.5\sigma$ of our estimates.
	Regarding $\lambda$ itself, the Fisher-Huse upper bound $\lambda \leq {5}/{4}$ (valid for the $2D$ Ising model)~\cite{fisherNonequilibriumDynamicsSpin1988} and the lower bound $\lambda\geq {d}/{2}=1$~\cite{fisherNonequilibriumDynamicsSpin1988,yeungBoundsDecayAutocorrelation1996} are all obeyed.
	The numerical precision is at least comparable to earlier simulations~\cite{christiansenAgingLongrangeIsing2020,lorenzNumericalTestsLocal2007,henkelTwotimeAutocorrelationFunction2004,menyhardDomaingrowthPropertiesTwodimensional1994} with the noise in $\plath$ being the primary source of error.
	If one asserts $z=2$, which can be done for the $2D$ Ising model without greater doubts, we may estimate $\lambda^* = 1.2534(34)$ at $T=0.2T_c$, which is our most reliable result, using the equilibrium plateau of the two-time autocorrelator.
	Our results are, thus, compatible with the stronger Fisher-Huse conjecture~\cite{fisherNonequilibriumDynamicsSpin1988} $\lambda = {5}/{4}$ in the $2D$ Ising model.

	We point out that our results do not support several earlier numerical and analytical results which have led to somewhat larger values of $\lambda$ and which do not obey the Fisher-Huse upper bound either \cite{vadakkayilFinitesizeScalingStudy2019,midyaAgingFerromagneticOrdering2014,liuNonequilibriumAutocorrelationsPhaseordering1991}.
	These are based on the large-$y$ asymptotics of $f_C(y)$.
	A systematic investigation on unrecognized biases might be in order.

	Thus, after testing and evaluating the method of quantifying aging via finite-size effects in the square lattice Ising system, we plan to extend it for more complicated models, e.g., by introducing long-range interaction, facilitated through our high-speed algorithm~\cite{mullerFastHierarchicalAdaptive2023}; or by introducing more spin values via the $q$-state Potts model.

	\ack{We thank Fabio M\"uller for helpful discussions.}

	\funding{This project was funded by the Deutsch-Franz\"osische Hochschule (DFH) / Universit\'e Franco-Allemande (UFA) under Grant No.~CDFA 02-07. D\,.W.~and W.\,J.~acknowledge further support by the Deutsche Forschungsgemeinschaft	(DFG, German Research Foundation) under Grant No.~560\,547\,547 (Project ID JA 483/36-1). M.\,H.~was supported by the French ANR-PRME UNIOPEN (ANR-22-CE30-0004-01).}
	
	\providecommand{\href}[2]{#2}

	\end{document}